\documentclass[11pt,notitlepage,tightenlines,eqsecnum,floats,aps,amsmath,superscriptaddress,amssymb,nofootinbib,longbibliography]{revtex4-1}

\usepackage{graphicx}
\usepackage{amsmath,amssymb}
\usepackage{hyperref}
\usepackage[dvipsnames]{color,xcolor,colortbl}
\usepackage[normalem]{ulem}
\usepackage{comment}
\usepackage{wrapfig}

\newcommand{\be}{\nopagebreak[3]\begin{equation}}
\newcommand{\ee}{\end{equation}}
\newcommand{\bea}{\begin{eqnarray}}
\newcommand{\eea}{\end{eqnarray}}

\begin{document}

\title{Vacuum-purified Hawking radiation from evaporating black holes:\\ Lessons from moving mirrors}

\author{Ivan Agullo}
\email{agullo@lsu.edu}
\affiliation{Department of Physics and Astronomy, Louisiana State University, Baton Rouge, LA 70803, U.S.A.
 }
\author{Paula Calizaya Cabrera}
\email{pcaliz1@lsu.edu}
\affiliation{Department of Physics and Astronomy, Louisiana State University, Baton Rouge, LA 70803, U.S.A.
 }%
\author{Beatriz Elizaga Navascués}%
\email{beatriz.elizaga@uclm.es}
\affiliation{Department of Physics, Faculty of Environmental Sciences and Biochemistry, University of Castilla - La Mancha, 45071 Toledo, Spain.
 }

\begin{abstract}

This article investigates the possibility that Hawking-like quanta emitted by a moving mirror can be purified by late-time vacuum fluctuations, as proposed in Hotta, Schützhold and Unruh [Phys. Rev. D 91,
124060 (2015)]. Our motivation originates from recent discussions in Wald [Phys. Rev. D 100, 065019
(2019)] and Osawa et al. [Phys. Rev. D 110, 025023 (2024)] on whether vacuum purification necessarily entails a prohibitively large (indirect) energy cost, and our goal is to help clarify this issue. We identify the aspects of the mirror trajectory that determine the partners of Hawking quanta, as well as those that govern the energy carried to future null infinity. This allows us to highlight a fundamental disconnection within quantum field theory between the fluxes of quantum information (or purification) and energy. Throughout, we focus on quantities such as local correlation functions and energy fluxes, thereby avoiding reliance on a particle-based interpretation. Finally, we introduce an analytic mirror trajectory that produces Hawking radiation with an adiabatically varying temperature, mimicking the emission from an evaporating black hole. Our analysis identifies constraints on the mirror trajectory under which vacuum purification remains compatible with a prescribed energy budget, and we discuss the lessons that may be drawn from this model for realistic evaporating black holes.
\end{abstract}

\maketitle

\section{\label{sec:1} Introduction}

A seemingly natural mechanism by which black holes could process information during their evaporation entails storing the  Hawking ``partners''---the field modes entangled with the radiation---until the black hole mass becomes Planckian. At that point, either the black hole becomes quantum mechanically unstable and releases the information in a final emission, or it stabilizes, in which case the information would remain stored inside the black hole remnant indefinitely. The remnant scenario comes with its own challenges \cite{Wald:2019ygd,Wald:1995yp,Unruh:2017uaw}. On the other hand, the release of information at the end of the evaporation process remains a plausible possibility \cite{Unruh:2017uaw,H_S_U_2015,Osawa:2024fqb,Wald:2019ygd,Carlitz:1986nh,Wilczek:1993jn}.

A potential problem with the information-release scenario arises from energy considerations. An enormous number of partner particles must be emitted—as many as the Hawking particles that escaped to infinity—by an object whose surface is Planckian in size. Given the system's Planck-scale dimensions, each partner presumably has a Planckian wavelength.\footnote{A portion of the community believes that this argument may be incorrect (see e.g. \cite{Ashtekar:2025ptw,Rovelli:2024sjl} and references therein), on the grounds that black holes with Planck-scale areas could still enclose enormous interior volumes.} However, this would lead to an energetically implausible situation: the total energy of the partners would vastly exceed not only the remaining energy of the system but even the mass of the progenitor star that originally collapsed to form the black hole.

Several years ago, Hotta, Schützhold, and Unruh \cite{H_S_U_2015} pointed out that this energy estimate might be incorrect by many orders of magnitude. To make their point explicit, they considered the well-known moving mirror analogy for Hawking radiation—a $1+1$ dimensional toy model first introduced by Davies and Fulling in the 1970s \cite{Fulling_Davies_1976,Fulling_Davies_1977}. In this model, an accelerating mirror following a particular trajectory in flat spacetime spontaneously emits Hawking-like radiation. This toy model captures many essential features of black hole evaporation while avoiding complications related to gravitational collapse and curvature singularities. Notably, evolution in this model allows one to trace how the outgoing radiation becomes purified at future infinity.

\begin{figure}
\centering   
    \includegraphics[width=.48\linewidth]{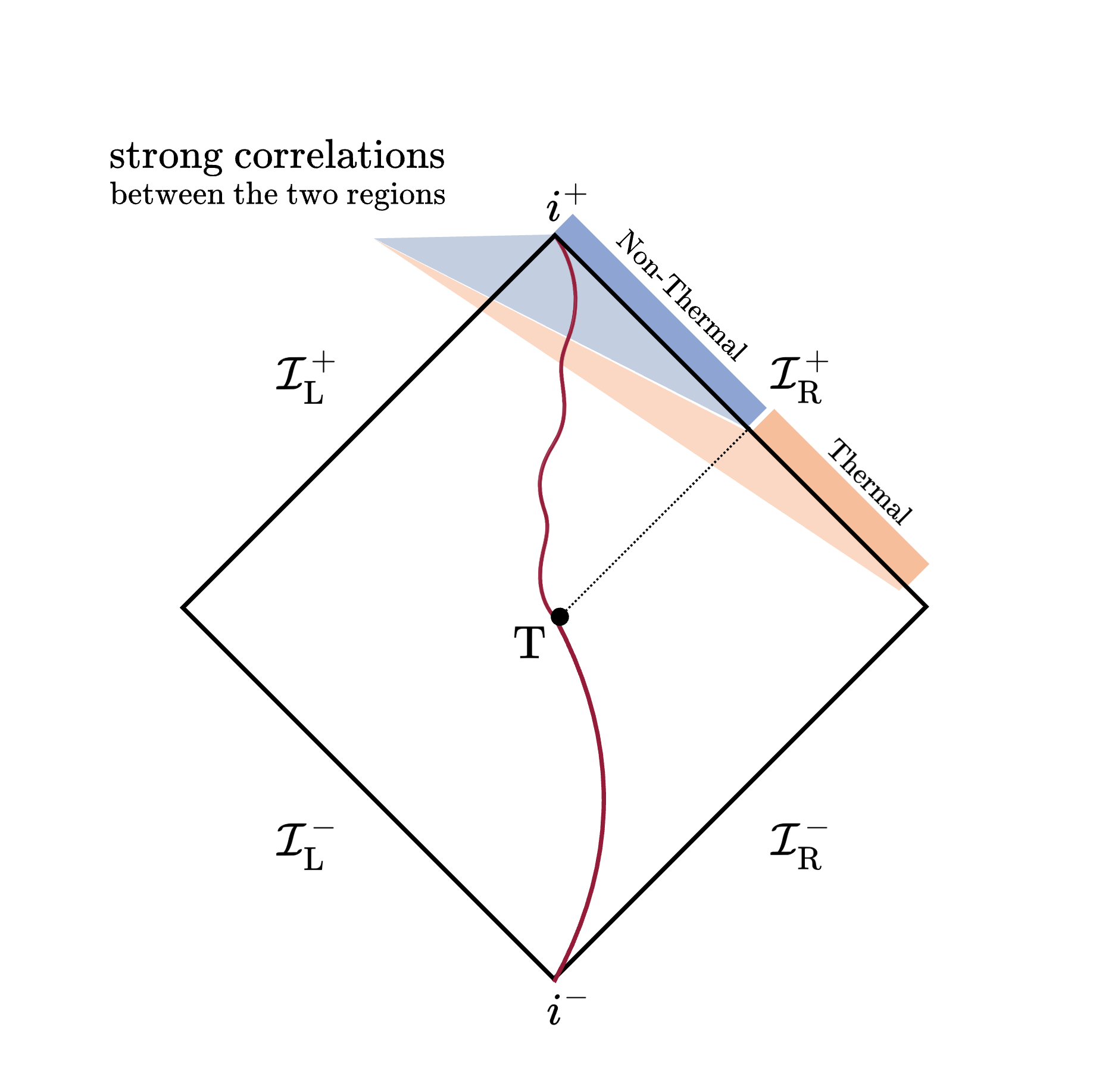}
    \includegraphics[width=.48\linewidth]{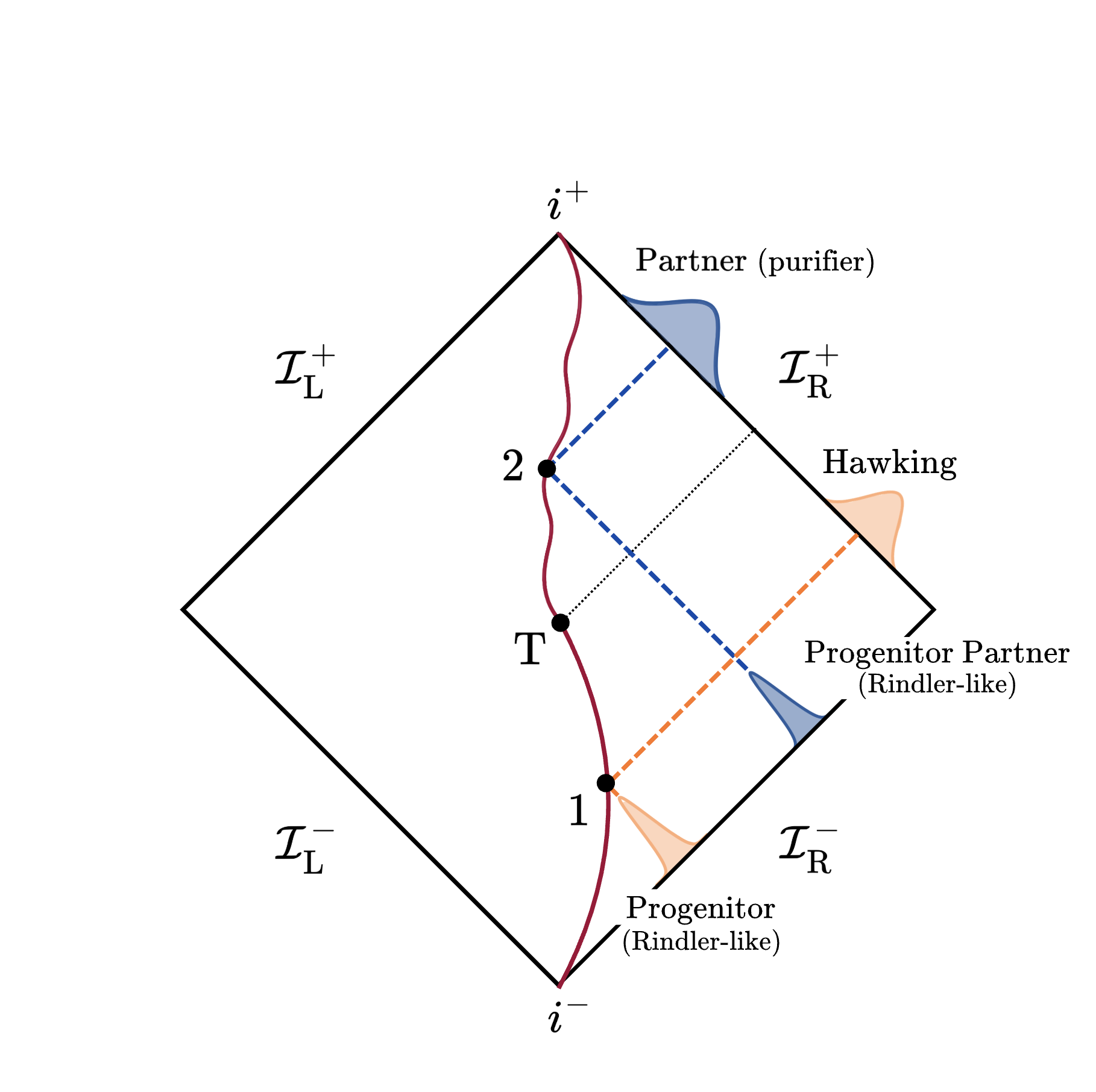}
      \caption{ Mirror trajectory beginning at $i^-$ and ending at $i^+$. The left panel depicts a mirror that initially emits thermal radiation to $\mathcal{I}^+_R$. At the point $T$, the trajectory changes and the emitted radiation ceases to be thermal. Strong correlations between widely separated points on $\mathcal{I}^+_R$ ensure that the total state of the field remains pure. The right panel shows the same mirror configuration as in the left panel, but focuses on an inertial particle arriving at $\mathcal{I}^+_R$ as a Hawking particle. The figure indicates the location of its progenitor mode on $\mathcal{I}^-_R$, as well as that of its partner mode. The location of the partner on $\mathcal{I}^-_R$ is identified by propagating the Hawking mode backwards in time and reflecting it across the asymptote of the mirror trajectory prior to point~$T$. The shape of the partner mode upon reaching $\mathcal{I}^+_R$, as well as its energy content, is determined by the state of motion of the mirror in the neighborhood of the reflection event (point~2).
}
      \label{fig2}
\end{figure}

Hotta, Schützhold, and Unruh showed that, if the mirror eventually returns to inertial motion, then the early thermal-like Hawking quanta can be purified at {\em no energy cost}. They argued that the purification is provided by local \emph{vacuum fluctuations}—that is, by field modes localized in regions where the state is indistinguishable from the vacuum. Early-time observers at future null infinity would interpret the state as thermal, while  any measurement by a local observer at late time would fail to detect deviations from the vacuum. 
Yet, strong global correlations between field modes supported at early and late times—inaccessible to either of the two local observers—render the full state pure; see Fig.~\ref{fig2} (the quantitative details supporting this statement are discussed later).

The idea that Hawking radiation can be purified at no (or negligible) energy cost is fascinating. It reveals that energy and quantum information do not necessarily go hand in hand  \cite{H_S_U_2015}. This is yet another manifestation of the rich structure of the quantum field-theoretic vacuum—it constitutes a vast reservoir of quantum degrees of freedom with which other field modes can become entangled. Whether such a mechanism plays a role in actual black hole evaporation remains to be established, but it certainly warns us that naive energy budget arguments may be misleading.

Given its potential relevance to the information loss paradox, the proposal by Hotta, Schützhold, and Unruh was further scrutinized by Wald in Ref.~\cite{Wald:2019ygd}, who provided a detailed analysis that clarified and refined some of the calculations. Wald, however, concluded that vacuum purification of Hawking-like radiation is not a free lunch. While it is true that partner modes can carry negligible energy, an indirect energy cost appears elsewhere in the system. This ``hidden'' energy cost brings the energetic concerns back into the discussion. 

The story does not end there. In a later work \cite{Osawa:2024fqb}, Osawa, Lin, Nambu, Hotta, and Chen contested the necessity of this hidden energy cost—in the way it was argued in Ref.~\cite{Wald:2019ygd}—by constructing a specific example in which the arguments of Ref.~\cite{Wald:2019ygd} do not materialize.

In view of the interest and potential consequences of this discussion, it would be desirable to identify the precise assumptions in Wald’s analysis that lead to the indirect energy cost claimed, and to explore whether they can be relaxed in physically realistic scenarios.  These are the main goals of the present article.

We begin with a pedagogical discussion of what determines the energy carried by partner modes to infinity, shedding light on why energy and information are not necessarily correlated. After identifying the assumption in Ref.~\cite{Wald:2019ygd} that leads to the conclusion of an indirect energy cost, we proceed to the second main contribution of this article: We introduce a mirror trajectory designed to produce radiation mimicking that emitted by a {\em slowly evaporating} black hole. The radiation temperature increases adiabatically, as it would in a real black hole. Once the temperature reaches a value of the order of the Planck scale, the system undergoes a short transition period, after which no further radiation is emitted, simulating the end of black hole evaporation.

This model is analytically tractable and carries interesting messages. In particular, we show that Hawking-like radiation can indeed be purified by vacuum fluctuations with very large wavelengths and carrying no energy.  Under reasonable choices of the mirror trajectory during the transition to inertial motion at the end of emission period, there is no prohibitive hidden energy cost.  Finally, we analyze the details of the purification process and discuss which insights from this toy model may, or may not, apply to realistic black hole scenarios.

Throughout this article we use units in which $\hbar=c=G=k_B=1$.

\section{\label{sec:2} Moving mirrors, energy fluxes and correlations}

The moving mirror analogue of Hawking radiation is commonly formulated within the simple setting of a scalar field $\varphi$ with zero rest mass propagating in a 1+1-dimensional flat spacetime. In this section, we summarize this model and show how, using tools from conformal field theory (CFT), one can determine under what conditions the mirror radiates, as well as how to compute the associated energy flux and the number of particles created by the mirror. While these results are not new (see e.g. Refs.~\cite{Fabbri:2004yy,Fabbri:2005mw}), we present them  in a pedagogical way for the convenience of readers who may be unfamiliar with the topic. 

\subsection{The classical theory}

Consider a real, massless scalar field \(\varphi\) propagating in \(1+1\)-dimensional Minkowski spacetime.  Its quantum theory is ill-defined if \(\varphi\) is interpreted as a  fundamental, physical field: the vacuum two-point distribution does not exist due to an infrared divergence. Nevertheless, the theory is well-defined when formulated in terms of the one-form \(F_a = \nabla_a \varphi\). From this perspective, \(\varphi\) plays the role of a gauge potential.

The equations of motion of the fundamental theory are
\begin{equation}
    \nabla^a F_a = 0\,, \quad \nabla^a\, {{}^*F_a} = 0\,,
\end{equation}
where \({}^*F_a\) denotes the Hodge dual of \(F_a\). These two equations can be written in differential form notation as \(d^*F = 0\) and \(dF = 0\), respectively. The latter implies that \(F_a = \nabla_a \varphi\), and then the former is equivalent to the Klein–Gordon equation for \(\varphi\), namely,
\begin{equation}
    \nabla^a \nabla_a \varphi = 0\,.
\end{equation}

The theory is gauge-invariant under transformations \(\varphi \to \varphi + \alpha\), for any spacetime function \(\alpha\) satisfying \(d\alpha = 0\). It is also invariant under duality transformations \(F_a \leftrightarrow {}^*F_a\), and under conformal coordinate transformations.\footnote{This theory is the one-form analog of sourceless electrodynamics, in which the electromagnetic field strength \(F_{ab}\) satisfies Maxwell’s equations \(dF = 0\) and \(d^*F = 0\). Maxwell theory is gauge-invariant and, in the absence of sources,  conformally invariant and duality-invariant.} In the CFT language, the components of \(F_a\) are referred to as conformal primary fields.

Let us now introduce the standard retarded and advanced null coordinates in terms of the inertial time \( t \) and space \( x \) coordinates as:
\begin{equation*}
    u = t - x , \quad v = t + x.
\end{equation*} 
In these null coordinates, the Klein–Gordon equation takes the form
\begin{equation}\label{KG}
    \partial_u \partial_v \varphi = 0.
\end{equation}
Any field configuration of the form \(\varphi(u,v) = f(u) + g(v)\), with \( f(u) \) and \( g(v) \) arbitrary (smooth) functions satisfies the equation of motion.

\subsection{The mirror}

In one spatial dimension, a perfect mirror for the field $\varphi$ can be modeled as a point in space where the field is forced to vanish. That this produces a mirror can be understood as follows.

Let $\gamma(\lambda) = (u(\lambda), v(\lambda))$ denote a timelike curve describing the mirror's trajectory  in spacetime. It is convenient to use the null coordinate \( u \) as the parameter of the curve, allowing us to write its worldline as \(\gamma(u) = (u, v = p(u))\); in this way, the mirror's trajectory is fully specified by a single function \( p(u) \), and it is timelike if and only if $\dot{p}>0$. The mirror imposes the boundary condition $ \varphi|_{\gamma} =0$. Equivalently:
\begin{equation} \label{mirrcond}
    \varphi(u, v = p(u)) = 0, \quad \forall u.
\end{equation}

The well-posedness of the initial value problem for this theory implies that a unique solution is determined by specifying initial data at past null infinity (see Fig.~\ref{fig2}). Consider, for instance, the initial data:
\[
\varphi|_{\mathcal{I}_L^-} = 0, \quad \varphi|_{\mathcal{I}_R^-} = e^{-i\omega v}.
\]
It corresponds to a left-moving plane wave emanating from the right portion of past null infinity (\( x \to \infty \) and \( t \to -\infty \), or equivalently \( u \to -\infty \)).

The solution of the Klein–Gordon equation with this initial data is
\begin{equation} \label{rfl}
    \varphi(u, v) = e^{-i\omega v} - e^{-i\omega p(u)},
\end{equation}
on the right-hand side of the mirror, and \( \varphi(u,v) = 0 \) on the left. To verify that this is indeed the desired solution, note that: (i) It is of the form \( f(u) + g(v) \), and therefore solves the Klein–Gordon equation \eqref{KG}; (ii) It satisfies the prescribed initial data; and (iii) It satisfies the boundary condition imposed by the mirror, $\varphi|_{\gamma} =0$. 

The solution \eqref{rfl} physically represents a left-moving plane wave approaching from \( x \to \infty \), which is reflected at the mirror into a right-moving wave that travels back to infinity ---the reflection also induces a relative phase equal to $\pi$.

Notice that the field degrees of freedom on the left and right sides of the mirror become disconnected, as the mirror condition \eqref{mirrcond} precludes the flow of information or energy from one side of the mirror to the other. Without loss of generality, from now on we focus on one side only, namely the region of spacetime to the right of the mirror.

The central question we aim to address is the following: if the field is prepared in the inertial vacuum state at past null infinity \( \mathcal{I}_R^- \), how is this state perceived at future null infinity \( \mathcal{I}_R^+ \) by inertial observers?

\subsection{Correlations, energy fluxes, and particles}

The fact that we are dealing with a 1+1-dimensional CFT makes it particularly simple to compute two-point correlations, energy fluxes, and particle numbers. The expressions reported here are all derived using well-known tools from CFT \cite{Fabbri:2005mw,Fabbri:2004yy,DiFrancesco:1997nk}. 

At ${\mathcal{I}_R^-}$ (corresponding to $u \to -\infty$), the relevant  primary field is $\partial_v\varphi$. The two-point distribution for the vacuum state $|{\rm in}\rangle$ there takes the standard vacuum form in a 1+1 CFT:
\begin{equation} \label{two-point-in}
    \left. \langle {\rm in} |
    \, \partial_v  \hat{\varphi}(v_1)\, \partial_v\hat{\varphi}(v_2) \,
    | {\rm in} \rangle \right|_{\mathcal{I}_R^-}
    = -\frac{1}{4\pi} \frac{1}{(v_1 - v_2 - i\varepsilon)^2},
\end{equation}
where, as usual, a limit $\varepsilon \to 0^+$ is understood. 

The relevant component of the (renormalized) stress-energy tensor at ${\mathcal{I}_R^-}$ is
\begin{equation}\label{Tvvscri-} 
    \left. \langle {\rm in} | \hat{T}_{vv} | {\rm in} \rangle \right|_{\mathcal{I}_R^-} = 0.
\end{equation}
The operator $\hat{T}_{vv}$ describes an incoming  energy flux  from ${\mathcal{I}_R^-}$, so Eqn.~\eqref{Tvvscri-} informs us that there is no incoming energy flux when the field is in the vacuum state $|{\rm in} \rangle$—as expected.

To propagate these quantities to ${\mathcal{I}_R^+}$, all we need is the function $p(u)$ describing the mirror's trajectory. This is evident from Eq.~\eqref{rfl}, which shows that a plane wave $e^{-i\omega v}$ at ${\mathcal{I}_R^-}$ becomes $e^{-i\omega p(u)}$ at ${\mathcal{I}_R^+}$. Since the vacuum at ${\mathcal{I}_R^-}$ is defined using plane-wave initial data, it is not surprising that knowledge of the function $p(u)$ is sufficient to evolve any vacuum expectation value to ${\mathcal{I}_R^+}$. 

The function $p(u)$ has a simple interpretation in terms of null geodesics that bounce off the mirror: it relates the arrival coordinate $u$ at ${\mathcal{I}_R^+}$ to the departure coordinate $v = p(u)$ at ${\mathcal{I}_R^-}$. It turns out that correlations and energy fluxes at ${\mathcal{I}_R^+}$ can be easily computed by interpreting the relation $v = p(u)$ as a change of coordinates, and applying the transformation laws for conformal primary fields and the energy-momentum tensor in CFT \cite{Fabbri:2005mw,Fabbri:2004yy,DiFrancesco:1997nk}. One obtains:
\begin{equation} \label{two-point-out}
    \left. \langle {\rm in} | 
    \partial_u\hat{\varphi}(u_1)\, \partial_u\hat{\varphi}(u_2) 
    | {\rm in} \rangle \right|_{\mathcal{I}_R^+}
    = -\frac{\dot{p}(u_1)\dot{p}(u_2)}{4 \pi [p(u_1)-p(u_2) - i\varepsilon]^2},
\end{equation}
and \cite{Fulling_Davies_1976}
\begin{equation} \label{Tuu}
    \left. \langle {\rm in} | \hat{T}_{uu} | {\rm in} \rangle \right|_{\mathcal{I}_R^+}
    = \frac{1}{16 \pi} \left[ \left( \frac{\ddot{p}}{\dot{p}} \right)^2 - \frac{2}{3} \frac{\dddot{p}}{\dot{p}} \right],
\end{equation}
The right-hand-side is proportional to the Schwarzian derivative of the function $p$ \cite{DiFrancesco:1997nk}. The expectation value $\langle {\rm in} | \hat{T}_{uu} | {\rm in} \rangle |_{\mathcal{I}_R^+}$ describes the flux of energy arriving at $\mathcal{I}_R^+$ when the field is prepared in the $ | {\rm in} \rangle$ vacuum at $\mathcal{I}_R^-$.

One can further exploit conformal invariance to easily determine the family of trajectories $v=p(u)$ that leave the vacuum invariant. These are determined by elements of the Lie subgroup $SO(2,2)$ of the conformal group in $1+1$ dimensions \cite{Fulling_Davies_1976}:
\begin{equation}
p(u) = \frac{a\, u + b}{c\, u + d}~,
\end{equation}
where $a, b, c, d \in \mathbb{R}$ and $ad - bc = 1$. In particular, spacetime translations correspond to $c=0$ and $a/d=1$, whereas dilations and boosts are obtained when $c=b=0$.  For these trajectories, it is straightforward to verify that:
\begin{equation}
    \left. \langle {\rm in} | 
    \partial_u\hat{\varphi}(u_1)\, \partial_u\hat{\varphi}(u_2) 
    | {\rm in} \rangle \right|_{\mathcal{I}_R^+}
    = -\frac{1}{4 \pi} \frac{1}{(u_1 - u_2 - i\varepsilon)^2},
\end{equation}
and
\begin{equation}
    \left. \langle {\rm in} | \hat{T}_{uu} | {\rm in} \rangle \right|_{\mathcal{I}_R^+} = 0.
\end{equation}
confirming that the state $|{\rm in} \rangle$ is perceived at $\mathcal{I}_R^+$ as vacuum---reflection at the mirror leaves the vacuum invariant.  These are the only trajectories with this property.

The trajectory proposed by Davies and Fulling to mimic Hawking radiation is \begin{equation}
 \label{tranjectoryFD} p(u) =- \kappa^{-1} e^{-\kappa u}\, .\end{equation} This trajectory does not belong to the conformal family discussed above. Using Eqns.~\eqref{Tuu} and \eqref{two-point-out}, one obtains instead:
\begin{equation}
    \left. \langle {\rm in} | \hat{T}_{uu} | {\rm in} \rangle \right|_{\mathcal{I}_R^+} = \frac{\kappa^2}{48 \pi},
\end{equation}
and
\begin{equation} \label{thermalcorrel}
    \left. \langle {\rm in} | 
    \partial_u \hat{\varphi}(u_1) \, \partial_u \hat{\varphi}(u_2) 
    | {\rm in} \rangle \right|_{\mathcal{I}_R^+}
    = - \frac{\kappa^2}{16\pi\,\sinh^2{\left[\kappa (u_1 - u_2 - i\varepsilon)/2\right]}}.
\end{equation}
These correspond, respectively, to a thermal flux of energy and a thermal two-point function\footnote{Thermality follows from the periodicity of the  correlation function under shifts of $u_1 - u_2$ in the imaginary axis, with period $2\pi \,i/\kappa$.}  at temperature $T = \kappa / (2\pi)$. Since the state $|{\rm in} \rangle$ is Gaussian, the two-point function fully determines the quantum state. Thus, Eqn.~\eqref{thermalcorrel} reveals that the vacuum state $ | {\rm in} \rangle$   is perceived at ${\mathcal{I}_R^+}$ as a thermal state by inertial observers. 

 In particular, the expected number of particles at ${\mathcal{I}_R^+}$ can be computed as follows. Let $W^{\rm out}_\omega(u)$ be a positive-frequency wavepacket on ${\mathcal{I}_R^+}$,  peaked around a central frequency $\omega$. The mean number of quanta in this mode is obtained by smearing the two-point function of the primary field with the wavepacket as follows:
\begin{equation}\label{Nout}
    \langle {\rm in} | \hat{N}_{\rm out}(\omega) | {\rm in} \rangle = 
    4 \int du_1 \, du_2 \, 
    W^{\rm out}_\omega(u_1) \, W^{\rm out *}_\omega(u_2) \,
    \langle {\rm in} | 
    \partial_u \hat{\varphi}(u_1)\, \partial_u \hat{\varphi}(u_2)
    | {\rm in} \rangle.
\end{equation}

This equation follows from the definition of the annihilation operator 
$\hat{a}_{\rm out}(\omega) = (W^{\rm out}_\omega, \hat{\varphi})_{\text{KG}}$, 
where the Klein-Gordon product of two functions $f$ and $g$ is given by
\[
(f, g)_{\text{KG}} = i \int du \, \left(\bar f \,{\partial}_u \, g - \bar g \,{\partial}_u \, f\right)
\]
on ${\mathcal{I}_R^+}$ (the bar denotes complex conjugation). Using integration by parts, one can move the derivatives to act only on the field operator.

In the limit of a sharp-frequency mode, $W^{\rm out}_\omega(u) = e^{-i\omega u} / \sqrt{4\pi\omega}$, equation \eqref{Nout} yields the familiar Bose-Einstein particle number spectrum:
\begin{equation}
    \langle {\rm in} | \hat{N}_{\rm out}(\omega) | {\rm in} \rangle 
    = \frac{1}{e^{2\pi\omega/\kappa} - 1}  \delta(0),
\end{equation}
where the divergence $\delta(0)$ arises from the non-normalizability of idealized plane waves. The use of properly normalized wavepackets regulates this mathematical divergence.

In this way, we reproduce the conclusion of Davies and Fulling: a mirror following the trajectory $p(u) = -\kappa^{-1} e^{-\kappa u}$ emits radiation as if it were a black body at temperature $\kappa/(2\pi)$.

If the mirror’s trajectory were  to remain of the form  $p(u) = -\kappa^{-1} e^{-\kappa u}$,  for all $u$, its inertial velocity would be 
\begin{equation}\label{velocitymirror}
\frac{dx}{dt}=\frac{\dot p-1}{\dot p+1} = \frac{1-e^{\kappa u}}{1+e^{\kappa u}}.
\end{equation}
As $u \to \infty$, this velocity approaches one—the speed of light—implying that the mirror would end up in $\mathcal{I}^+_L$ rather than at $i^+$ (see Fig.~\ref{fig2}). For such a trajectory the field evolution from $\mathcal{I}^-_R$ to $\mathcal{I}^+_R$ is not one-to-one.  To ensure that all information flows exclusively from $\mathcal{I}^-_R$ to $\mathcal{I}^+_R$, the mirror trajectory must be modified so that it begins at $i^-$ and ends at $i^+$. We will focus on this family of trajectories in the remainder of this article.

\section{\label{sec:3} Partner modes and their energy}

This section is divided into two parts. In the first part, in order to build intuition, we present a heuristic description of partner modes —defined as field modes that purify the Hawking-like radiation emitted by the mirror \cite{H_S_U_2015}—and analyze what determines the energy they carry to ${\mathcal{I}_R^+}$. We argue that the energy carried by the partner modes  can be modified at will by adjusting the mirror's trajectory at late times, without affecting the early thermal emission and the fact that the partner mode purifies a previously emitted Hawking particle.  In the second part, following Ref.~\cite{Agullo:2025dxp}, we summarize the formal definition of partner modes in quantum field theory, their existence and uniqueness, and the method to compute them in general settings. Although some of the tools introduced in that subsection are not strictly necessary for this article —since, in the comparatively simple context of moving mirrors, we can employ shortcuts to compute partner modes— we believe it is conceptually important for the discussion of purification of Hawking-like radiation to make  the foundation for the notion of partner modes available.

\subsection{The energy of partners: intuitive understanding}

Consider the trajectory depicted in Fig.~\ref{fig2}, in which a mirror initially follows the trajectory \( p(u) = -\kappa^{-1} e^{-\kappa u} \)—thus emitting Hawking radiation at temperature \( \kappa/(2\pi) \)—and then transitions to a different trajectory, which we leave unspecified for the moment.

The way the \( |{\rm in}\rangle \) vacuum state is perceived at \( \mathcal{I}_R^+ \) can be analyzed in two complementary ways.

The first strategy focuses on the particle content of the state at \( \mathcal{I}_R^+ \). This requires adopting the natural notion of particle at \( \mathcal{I}_R^+ \)—namely, the one selected by invariance under translations in the retarded time $u$ along \( \mathcal{I}_R^+ \)—and investigating the number of particles contained in suitably chosen field modes, such as wavepackets. This calculation provides a particle interpretation of the \( |{\rm in}\rangle \) state at  \( \mathcal{I}_R^+ \), informing us how much it deviates from the natural vacuum in that asymptotic region.

The second strategy employs quantities such as the energy flux \( T_{uu} \) at \( \mathcal{I}_R^+ \) and two-point correlations of the form
\[
\langle {\rm in} | \partial_u \hat{\varphi}(u_1) \, \partial_u \hat{\varphi}(u_2) | {\rm in} \rangle \big|_{\mathcal{I}_R^+}.
\]
These quantities are invariant in the sense that they do not depend on any specific choice of modes, and hence reflect general features of the quantum state as perceived by any inertial observer. 

The first strategy is illustrated in the right panel of Fig.~\ref{fig2}. A wavepacket representing a particle is shown at early times \( u \) in \( \mathcal{I}_R^+ \). The support of this wavepacket is concentrated in the region where the mirror emits thermal radiation.\footnote{Strictly speaking, a wavepacket constructed only from positive frequencies cannot have compact support. However, its amplitude decays very rapidly outside a region of width \( \Delta u \approx (\Delta \omega)^{-1} \), where \( \Delta \omega \) denotes its bandwidth.} This mode of the field is thermally populated on the state \( |{\rm in}\rangle \), and its reduced quantum state is therefore mixed. The partner of this mode is another field mode with support in \( \mathcal{I}_R^+ \), which purifies the system---i.e., the joint two-mode state describing  the mode and its partner is pure (see next subsection for further details).

The partner mode is most easily identified by propagating the field mode backwards in time to \( \mathcal{I}_R^- \) (see Fig.~\ref{fig2}). There, as shown in Ref.~\cite{Wald:1975} and further discussed in Sec.~\ref{sec:6}, the mode takes the form of a Rindler mode  (specifically, a left-Rindler mode propagated to $\mathcal{I}_R^-$). This is fortunate, because from the analysis of the Unruh effect~\cite{PhysRevD.14.870}, we know that the partner of a Rindler mode in \( \mathcal{I}_R^- \) is obtained by reflecting the mode across the asymptote of the trajectory \( p(u) = -\kappa^{-1} e^{-\kappa u} \). The wave packet obtained by this mirror reflection defines the partner mode.

The well-posedness of the Cauchy problem ensures that the state of the Hawking mode plus its partner remains pure throughout their evolution. Therefore, the time-evolved partner mode must still purify the time-evolved Hawking mode. This permits us to locate the support of partner modes in \( \mathcal{I}_R^+ \) (see Fig.~\ref{fig2}).

In addition to locating the support, we are interested in determining the specific form of the partner when it reaches \( \mathcal{I}_R^+ \). This form is determined solely by the mirror’s trajectory at the moment of reflection, since the mode propagates freely both before and after the reflection. By modifying the mirror’s late-time trajectory, one can change the shape with which the partner mode arrives at \( \mathcal{I}_R^+ \). Importantly, irrespective of the shape the partner mode acquires after reflection, it continues to purify the same Hawking particle.

For instance, if the partner reflects off the mirror while it follows an inertial trajectory, the reflection may Doppler-shift the mode’s frequency content, but otherwise preserve its shape. As a result, the partner would reach \( \mathcal{I}_R^+ \) as a Doppler-shifted  Rindler mode. In Minkowski spacetime, Rindler modes carry no energy and correspond to vacuum fluctuations from the perspective of inertial observers. Therefore, when the mirror becomes inertial at late times, Hawking particles are purified by vacuum fluctuations. More generally, the specific form and particle content of the partner can be modified at will by changing the mirror’s trajectory—without affecting its purifying role.

A caveat to this argument is that the wavepackets describing both Hawking particles and their partners are not, strictly  speaking, compactly supported; they have highly suppressed tails extending over all of \( \mathcal{I}_R^+ \). These tails are typically neglected in the literature~\cite{Hawking_1975,Wald:1975,Wald:2019ygd}. Nevertheless,  it is important to keep in mind that all statements made above about the spatial support of partner modes involve an approximation.

On the other hand, rigorous local statements can be made using the aforementioned second strategy, which focuses on local correlations and  energy fluxes  rather than particle content. This is illustrated in the left panel of Fig.~\ref{fig2}. Initially, the mirror emits a thermal flux, and the two-point correlations in this region are given by Eq.~\eqref{thermalcorrel}. After the transition, two-point correlations are described by Eq.~\eqref{two-point-out} and are determined solely by the mirror’s trajectory {\em after} the transition. Similarly, the energy emitted after the transition depends only on the mirror’s later trajectory. In particular, if the trajectory becomes inertial at late times, then there is no outgoing energy flux. Nonetheless, there exist strong correlations between pairs of points in \( \mathcal{I}_R^+ \), one located before and one after the transition. These correlations are stronger than those that would exist in the natural vacuum at \( \mathcal{I}_R^+ \) (the so-called \( |{\rm out}\rangle \) vacuum)---see  the discussion around Eqn.~\eqref{corrinvsout} for further details. These enhanced correlations are responsible for restoring the purity of the \( |{\rm in}\rangle \) vacuum, even if this state appears thermal to observers localized at early times in \( \mathcal{I}_R^+ \).  It is a remarkable feature of quantum mechanics that quantum degrees of freedom carrying no energy can nevertheless be correlated and even entangled with other degrees of freedom.

\subsection{Modes and partners: a brief recap}\label{partner}
This subsection contains a brief recap of the definition and main properties of modes and their partners in quantum field theory, following the presentation in Ref.~\cite{Agullo:2025dxp} (see \cite{Botero:2003wov,H_S_U_2015,Trevison:2018ear} for earlier work). The calculation of partners of Hawking modes for the particular case of a mirror following the trajectory~\eqref{tranjectoryFD} is especially simple and does not require the general machinery summarized in this subsection. This is because the calculation reduces to identifying the partner of a Rindler mode in Minkowski spacetime—a well-known procedure~\cite{Wald:1975,Wald:1995yp}. 

Nevertheless, it is useful to provide a precise and general definition of the concept of a partner mode and its invariant properties, as well as general tools to compute them for arbitrary mirror trajectories and in other linear field theories.

Consider a Klein–Gordon field on a globally hyperbolic spacetime. We begin by defining a subsystem of this theory containing a single degree of freedom. Let \( f_A \) be a complex solution to the Klein–Gordon equation, restricted to those satisfying the unit Klein–Gordon norm condition \( (f_A, f_A)_{\rm KG} = 1 \). In the classical theory, such a solution defines a one–mode subsystem as follows: the linear span of the real and imaginary parts of \( f_A \) forms a two–dimensional symplectic subspace of the classical covariant phase space. This symplectic subspace constitutes the classical state space of a single degree of freedom of the field theory, defining a single ``mode of vibration'' of the classical system.

The corresponding quantum subsystem is defined as follows. From \( f_A \), we define a non-Hermitian operator as 
\begin{equation}
    \hat{a}_A = (f_A, \hat{\Phi})_{\rm KG} \, ,
\end{equation}
where \( \hat{\Phi} \) denotes the Klein--Gordon field operator-valued distribution. The  commutation relations for the field imply
\[
[\hat{a}_A, \hat{a}_A^\dagger] = (f_A, f_A)_{\rm KG} = 1 \, .
\]
Thus, \( \hat{a}_A \) and its adjoint satisfy the usual commutation relations of creation and annihilation operators—though note that, because \( f_A \) is arbitrary, \( \hat{a}_A \) does not necessarily annihilate a given Fock vacuum for the field.

The operator \( \hat{a}_A \) defines a single-mode subsystem of the field theory via the operator algebra generated by \( \hat{a}_A \) and \( \hat{a}_A^\dagger \). This algebra is isomorphic to that of a quantum harmonic oscillator, with the Hermitian and anti-Hermitian parts of \( \hat{a}_A \) playing the roles of \( \hat{x} \) and \( i\,\hat{p} \), respectively.

Note that the single-mode subsystem is defined by the algebra generated by \( \hat{a}_A \) and \( \hat{a}_A^\dagger \). Hence, we would obtain the same subsystem if we replaced \( \hat{a}_A \) by \( c_1 \hat{a}_A + c_2 \hat{a}_A^\dagger \), where \( c_1 \) and \( c_2 \) are complex numbers satisfying \( |c_1|^2 - |c_2|^2 = 1 \). Equivalently, \( f_A \) and \( \bar{c}_1 f_A - \bar{c}_2 \bar{f}_A \) define the same subsystem (where the bar denotes complex conjugation). Since this corresponds to a symplectic transformation of \( f_A \) and \( \bar{f}_A \), one says that single-mode subsystems are invariant under ``subsystem-local'' symplectic transformations. We thus identify each single-mode subsystem \( A \) with an equivalence class \( \{f_A\} \), where the equivalence relation is defined by subsystem-local symplectic transformations.

Suppose now that the field is prepared in a pure state \( |0\rangle \). For the purposes of this article, it suffices to restrict to \( |0\rangle \) being a Fock vacuum—i.e., a Gaussian state with zero mean, \( \langle 0 | \hat{\Phi} | 0 \rangle = 0 \).\footnote{Non-zero values of the first moments do not affect the discussion below. In particular, the partner of a given mode is insensitive to the value of the first moments.} We can obtain a reduced state for the single-mode subsystem \( A \) from \( |0\rangle \) by considering its algebraic restriction to \( A \). Heuristically, this can be understood as a partial trace over all field degrees of freedom except those belonging to \( A \).

In practice, this reduced state can be obtained straightforwardly by noting that, since \( |0\rangle \) is a Gaussian state, the reduced state is also Gaussian. As a result, it is completely and uniquely characterized by the expectation values \( \langle 0 | \hat{a}_A | 0 \rangle \) and \( \langle 0 | \hat{a}_A^{\dagger} | 0 \rangle \), both of which vanish, together with the expectation values of products of two such operators (i.e., the second moments). Any second moment can be decomposed into its symmetric and antisymmetric parts. The antisymmetric part corresponds to commutators of \( \hat{a}_A \) and \( \hat{a}_A^{\dagger} \), which are state-independent. Thus, it suffices to focus on the symmetrized second moments:
\begin{equation}
\sigma_A =
\begin{pmatrix}
\langle 0 | \{ \hat{a}_A, \hat{a}_A \} | 0 \rangle & \langle 0 | \{ \hat{a}_A, \hat{a}_A^\dagger \} | 0 \rangle \\
\langle 0 | \{ \hat{a}_A^\dagger, \hat{a}_A \} | 0 \rangle & \langle 0 | \{ \hat{a}_A^\dagger, \hat{a}_A^\dagger \} | 0 \rangle
\end{pmatrix} \, ,
\end{equation}
where \( \{ \cdot, \cdot \} \) denotes the anticommutator. The matrix \( \sigma_A \) is the \emph{covariance matrix} of the reduced state and encodes all physical information about the subsystem \( A \). Although one can write an explicit expression for the reduced density matrix \( \hat{\rho}^{\mathrm{red}}_A \) in terms of \( \sigma_A \), in practice most physical quantities can be directly extracted from \( \sigma_A \).
We are interested in invariant properties of the reduced state that do not depend on a particular representative \( f_A \) of the equivalence class \( \{ f_A \} \) defining the subsystem. More precisely, we focus on \emph{local symplectic invariants}—quantities that depend only on the subsystem itself, and not on any specific choice of basis. An example of such an invariant is the purity of the reduced state \( \hat{\rho}^{\mathrm{red}}_A \), which is given by \( 1 / \det \sigma_A \). This quantity is invariant under subsystem-local symplectic transformations, as such transformations preserve the determinant. Hence, the reduced state \( \hat{\rho}^{\mathrm{red}}_A \) is pure if and only if \( \det \sigma_A = 1 \).

As an illustration, consider a solution \( f_A \) that, at \( \mathcal{I}^{-} \), consists solely of positive frequencies with respect to \( v \)—for instance, a wavepacket. The subsystem it defines is then necessarily pure when the field is in the natural vacuum at \( \mathcal{I}^{-} \), which we have been calling \( | {\rm in} \rangle \). However, such a solution \( f_A \) cannot be compactly supported on \( \mathcal{I}^{-} \), since any compactly supported function necessarily contains negative-frequency components.

Conversely, if \( f_A \) is compactly supported at \( \mathcal{I}^{-} \), the single-mode subsystem it defines must be in a mixed state when the field is prepared in the vacuum state \( | {\rm in} \rangle \). This is because the condition of purity, \( \det \sigma_A = 1 \), of the reduced state is only satisfied if the equivalence class $\{f_A\}$ contains a purely positive-frequency mode \cite{Evaporation_2024}. 

Whenever the reduced state \( \hat{\rho}^{\mathrm{red}}_A \) is mixed and the global state \( |0\rangle \) is pure, the subsystem \( A \) must be entangled with the rest of the field modes. However, this entanglement may be distributed across infinitely many degrees of freedom, making it difficult to characterize. This is where the concept of a \emph{partner mode} becomes useful. If the global state is pure and Gaussian, one can always find another single-mode subsystem \( A_P \), independent of \( A \), that purifies it—that is, the joint reduced state of \( A \) and \( A_P \) is pure~\cite{Botero:2003wov,H_S_U_2015,Trevison:2018ear}. In this case, all correlations and entanglement between \( A \) and the rest of the field are encoded in \( A_P \). The partner subsystem \( A_P \) exists and is unique for linear field theories in Gaussian states. The notion of a partner can be generalized to situations in which the field is prepared in a mixed Gaussian state~\cite{Agullo:2025dxp}, but this generalization will not be needed in this article.

Given our definition of single-mode subsystems, the partner corresponds to an equivalence class $\{f_{A_P}\}$, and a representative $f_{A_P}$ can be computed as~\cite{Agullo:2025dxp}
\begin{equation} \label{partnerformula}
f_{A_P} = N\, \Pi^{\perp}_A(J \bar{f}_A) \, ,
\end{equation}
where $N=1/\sqrt{\det \sigma_A -1}$ is a normalization factor, and $\Pi^{\perp}_A = 1 - \Pi_A$ is the  projection operator onto the symplectic-orthogonal complement of the subsystem $A$, where
\[
\Pi_A = f_A (f_A, \cdot)_{\rm KG} - \bar{f}_A (\bar{f}_A, \cdot)_{\rm KG} \, .
\]
In Eqn.~\eqref{partnerformula}, $J$ is the complex structure associated with the vacuum state $|0\rangle$ \cite{Ashtekar:1975zn}\footnote{The complex structure $J$ associated with a vacuum state $|0\rangle$ is a real linear map on the space of solutions to the Klein--Gordon equation, whose square is minus the identity. It can be constructed from the symmetrized two-point function $W(x,x')=\langle 0| \{\hat\Phi(x),\hat\Phi(x')\}|0\rangle$ of the field operator $\hat{\Phi}(x)$ together with the symplectic structure (see Refs.~\cite{Ashtekar:1975zn,Agullo:2025dxp} for further details). Explicitly, the action of $J$ on any solution $f$ is given by
\[
Jf(x)=\int_{\Sigma} d\Sigma'^a 
\Big[f(x')\,\nabla'_aW(x,x')
- W(x,x')\,\nabla'_a f(x')\Big],
\]
where the integration is performed over primed coordinates on the Cauchy surface $\Sigma$ with oriented volume element $d\Sigma^a$, and the prime on $\nabla'_a$ indicates differentiation with respect to $x'$. The result of this integral is independent of the choice of $\Sigma$.}

The partner of a given mode conveys valuable information: In particular, it identifies the region of spacetime where the correlations and entanglement associated with that mode are localized.

In the context of moving mirrors, one is interested in the partners of Hawking modes. Knowledge of their localization is key to understand the potential  energy cost of purification: Do the partners carry significant energy to $\mathcal{I}_R^+$?

\section{Problems with vacuum purification?}\label{sec:4} 

That early-emitted Hawking quanta can be purified by field modes carrying no energy is a fascinating property of quantum field theory. Local vacuum fluctuations provide a vast reservoir of degrees of freedom which can become entangled with and purify real particles  in a different region of $\mathcal{I}^+$.  The Unruh effect reveals that vacuum fluctuations in one region of space are entangled with, and can purify, vacuum fluctuations in other regions. The crucial difference in the case of moving mirrors is that local vacuum fluctuations can purify {\em real particles}, rather than merely other vacuum fluctuations.

Wald suggested in Ref.~\cite{Wald:2019ygd} that, in the case of Hawking radiation emitted by a moving mirror, vacuum purification may not be completely cost-free in energetic terms. Rather, an indirect, highly energetic burst of radiation might appear elsewhere in $\mathcal{I}^+_R$. The argument goes as follows.

As discussed in the previous section, if the mirror returns to inertial motion at late times, the early-emitted Hawking particles can be purified by late-time Rindler modes in $\mathcal{I}^+_R$. This entanglement makes it evident that the $|{\rm in}\rangle$ state in which the field is prepared is quite different from the natural vacuum at $\mathcal{I}^+_R$---the $|{\rm out}\rangle$ vacuum---since in the  $|{\rm out}\rangle$ state  the late Rindler modes would be entangled with early Rindler modes rather than with positive frequency modes describing real particles. We refer to these early-time  Rindler modes as the “would-be partner of the partner”, as they are the modes with which the partners of Hawking quanta would be entangled if the quantum state were $|{\rm out}\rangle$.

Wald’s observation was that, since the would-be partner of the partner is not entangled with a late-time Rindler mode—because such a mode is instead entangled with a Hawking particle—the would-be partner of the partner cannot be in a state that  can be locally interpreted as vacuum. Wald concluded that there must exist particles in the region where the would-be partner of the partner is supported, and that the energy of such particles constitutes an indirect energetic cost to the purification of Hawking quanta. 

This is, indeed, an interesting observation. However, Osawa, Lin, Nambu, Hotta, and Chen argued in Ref.~\cite{Osawa:2024fqb} that the burst of energy discussed in \cite{Wald:2019ygd} is unrelated to vacuum purification, and hence should not be interpreted as an indirect energetic cost of it. They supported their argument by providing a counterexample: a mirror trajectory for which no energy burst appears at the location of the would-be partners of the partners. Instead, in their example, an energy burst arises at later times, after the purification zone at $\mathcal{I}_R^+$, and is attributed to a sharp deceleration of the mirror rather than to the purification mechanism.

Our contribution to this discussion is to highlight an assumption made in Ref.~\cite{Wald:2019ygd} that is not generally satisfied—namely, that the Rindler modes describing the would-be partners of the partners are independent of the field modes describing the Hawking particles; in other words, that they are orthogonal with respect to the Klein–Gordon product.  [Specifically, independence between the Hawking modes and the would-be partners  was used  in Ref.~\cite{Wald:2019ygd} around Eq.~(26).]  If this independence does not hold, then the energy cost associated with vacuum purification can already be included in the energy carried by the Hawking quanta themselves. In that case, there is no additional energy burst beyond the standard Hawking energy flux.

In the following section, we introduce a mirror trajectory that mimics an evaporating black hole, in the sense that the emitted radiation is approximately thermal with a temperature that increases adiabatically, as it would for an actual evaporating black hole. We further show that, in such a scenario, there is no indirect energy cost—indeed, there is no energy burst at all—and we verify that the Hawking quanta are not independent from the would-be partners of the partners.

\section{\label{sec:5} Modeling an evaporating black hole}

Previous discussions on purification through late-time vacuum fluctuations have focused on mirror trajectories $p(u)$ identical to the one considered by Davies and Fulling \cite{Fulling_Davies_1977} during an interval $u_0\leq  u < u_m$, followed by a transition period and, finally, inertial motion \cite{H_S_U_2015,Wald:2019ygd}:
\begin{align}\label{mirrorevp}
p(u)=
\begin{cases}
    p_{\mathrm{th}}(u)=v_H-\kappa^{-1} e^{-\kappa (u-u_0)}, & u_0\leq  u < u_{m},\\[6pt]
    p_{\mathrm{tr}}(u), & u \in [u_m,u_f),\\[6pt]
    p_{\mathrm{in}}(u), & u \ge u_f,
\end{cases}
\end{align}
where $p_{\mathrm{in}}(u)$ is a linear function of $u$.  
In the Hawking-radiation zone of $\mathcal{I}^+_R$---that is, for $u < u_m$---the trajectory $p_{\mathrm{th}}(u)$ produces an exactly thermal energy flux with temperature $T = \kappa / (2\pi)$. The constant $v_H$ determines the asymptote of $p_{\mathrm{th}}(u)$ and plays an important role in locating the partner modes, as further discussed below. For actual black holes, $v_H$ determines the location in $\mathcal{I}^-$ of the null ray that generates the event horizon.

The goal of this section is to replace $p_{\mathrm{th}}(u)$ with a trajectory $p_{\mathrm{evp}}(u)$ that produces radiation mimicking that of an actual evaporating black hole, i.e., with a slowly increasing temperature as the black hole loses mass.

The Hawking temperature of a spherically symmetric black hole is given by
\begin{equation}
    T_H = \frac{1}{8\pi M}~,
\end{equation}
where $M$ denotes the mass of the black hole. Hawking evaporation causes the mass of the system perceived in $\mathcal{I}^+$ to vary as
\begin{equation}\label{M}
    \dot{M}(u) = -\alpha M(u)^{-2},
\end{equation}
which integrates to
\begin{equation}\label{Msol}
    M(u) = M_0 \left[1 - 3\alpha M_0^{-3} (u - u_0)\right]^{1/3}.
\end{equation}
Here, $M_0=M(u_0)$ and $\alpha$ is a constant that depends on the number and spin of the radiated fields. Page computed $\alpha$ to be of order $10^{-4}$ in Planck units \cite{Page:1976df,Page:2013dx}, with massless particles---photons and gravitons---being the dominant channels of emission. In our analysis, instead, we will fix $\alpha$ to ensure consistency between the rate of change of $M(u)$ and the energy flux arriving at $\mathcal{I}^+$.

The mass loss described by Eqn.~\eqref{Msol} is very slow until $M(u)$ approaches the Planck scale. In this adiabatic regime, the radiation is approximately thermal, with an adiabatically varying temperature
\begin{equation}
    T_{\mathrm{evp}}(u) = \frac{1}{8\pi M(u)}.
\end{equation}
Our goal is therefore to propose a mirror trajectory that produces radiation with this temperature.
\subsubsection{Mirror trajectory}

In Ref.~\cite{Evaporation_2024}, we discussed the most general family of functions $p_{\rm evp}(u)$ (understood as ray-tracing relations in the context of black holes) with the desired properties. Here, we analyze one such trajectory, chosen for its simplicity.

A key observation is that the local temperature on $\mathcal{I}^+_R$ produced by a mirror trajectory $p(u)$ is given by $-\ddot{p}(u)/[2\pi\dot{p}(u)]$, if this combination of derivatives is adiabatic (see, e.g., \cite{Barcel2011} for a clear explanation). Thus, a temperature $T_{\rm evp}(u) = [8\pi M(u)]^{-1}$ with $M(u)$ as in Eqn.~\eqref{M} can be obtained demanding that
\begin{equation}\label{pdde0}
    \frac{\ddot{p}(u)}{\dot{p}(u)} = -\frac{1}{4M(u)}~.
\end{equation}
This is a second-order differential equation for $p(u)$, which can be solved on an interval $[u_0, u_m)$ in terms of two integration constants. These constants correspond to the values of $p(u)$ and its first derivative at an arbitrary instant. They can be identified with the freedom in choosing an affine parameter $v$ on $\mathcal{I}^{-}_R$. The properties of the radiation at $\mathcal{I}^{+}_R$, however, remain unaffected by this freedom, so the choice is inconsequential for our purposes.

The integration of Eq.~\eqref{pdde0} can be performed in two steps. First, we use the mathematical identity, valid for any sufficiently smooth function $p(u)$:
\begin{equation}\label{dotp}
    \dot{p}(u) = \dot{p}_0 \, \exp{\!\left[\int_{u_0}^{u} du'\, \frac{\ddot{p}(u')}{\dot{p}(u')}\right]},
\end{equation}
where $\dot{p}_0>0$ is a constant. Substituting Eqn.~\eqref{pdde0} into Eqn.~\eqref{dotp} and noting that, for $M(u)$ given in Eqn.~\eqref{Msol},  $\int du\, [4M(u)]^{-1} = -M^2(u)/(8\alpha)$, one obtains
\begin{align}\label{pdotexact}
    \dot{p}_{\rm evp}(u) = \dot{p}_0 \, \exp{\!\left[\frac{M^2(u) - M_0^2}{8\alpha}\right]}.
\end{align}

This expression shows that $\dot{p}(u)$ decreases exponentially as $M(u)$ decreases. Physically, this implies that the local redshift factor between $\mathcal{I}^{-}_R$ and $\mathcal{I}^{+}_R$ increases exponentially as $M(u)$ decreases.

Finally, integrating Eqn.~\eqref{pdotexact}, we obtain
\begin{align}\label{pexact}
p_{\rm evp}(u) = p_0 
+ 4\dot{p}_0\, e^{-M_0^2/(8\alpha)} 
\bigg\lbrace 
    & M_0\, e^{M_0^2/(8\alpha)}
    - M(u)\, e^{M^2(u)/(8\alpha)} \nonumber\\
    & + \sqrt{2\pi\alpha}\,
    \bigg[\mathrm{erfi}\!\left(\frac{M(u)}{\sqrt{8\alpha}}\right)
    - \mathrm{erfi}\!\left(\frac{M_0}{\sqrt{8\alpha}}\right)\bigg]
\bigg\rbrace,
\end{align}
where $p_0=p_{\rm evp}(u_0)$ is a constant, and $\mathrm{erfi}(x)$ is the imaginary error function, defined as
\begin{equation}
    \mathrm{erfi}(x) = \frac{2}{\sqrt{\pi}} \int_{0}^{x} dy\, e^{y^2}.
\end{equation}
For large real arguments, this function behaves as 
\begin{equation}
    \mathrm{erfi}(x) = \pi^{-1/2}\, e^{x^2}\, \big[x^{-1} + \mathcal{O}(x^{-3})\big].
\end{equation}
Eqn.~\eqref{pexact} provides the desired replacement of $p_{\rm th}(u)$. The interesting feature of this mirror trajectory is that, around any given instant $u_\star$ in the radiation zone $[u_0, u_m)$, it approximates an exponential trajectory that produces  thermal radiation with temperature $1/(8\pi M_\star)$, where $M_\star=M(u_\star)$ is the value of the function $M(u)$ in Eq.~\eqref{Msol} evaluated at $u = u_\star$. 

To see this explicitly, one expands $M(u)$ in an interval $u_\star \pm \Delta u$ centered at $u = u_\star\in [u_0, u_m)$. If $\Delta u$ is much smaller than $M_\star^2/\sqrt{\alpha}$ (yet still large if $M_0$ is greater than the Planck mass), the trajectory \eqref{pdotexact} is well approximated within the interval $u_\star \pm \Delta u$ by (next-to-leading-order corrections were reported in Ref.~\cite{Evaporation_2024}) 
\begin{equation} \label{pexpapprox}
    p_{\rm evp}(u) \approx v_\star^{(H)} - 4 M_\star \, \dot{p}_\star \, e^{-\frac{(u - u_\star)}{4 M_\star}},
\end{equation}
where the subscript $\star$ indicates evaluation at $u = u_\star$, and the constant $v_{\star}^{(H)}$ is given by
\begin{align} \label{vhinst}
v_{\star}^{(H)} = p_{\star} + 4 M_{\star} \dot{p}_{\star}.
\end{align}
This `constant', whose value varies with $u_\star$, was introduced in the context of evaporating black holes in  Ref.~\cite{Evaporation_2024}  as the location of the \emph{instantaneous would-be horizon}, since it corresponds to the position of a hypothetical event horizon that would form if $M(u)$ were forced to remain constant and equal to $M_\star$. The value of $v_{\star}^{(H)}$ plays a key role in determining the position of the Hawking partners. 

\begin{figure}
\centering   
    \includegraphics[width=.99\linewidth]{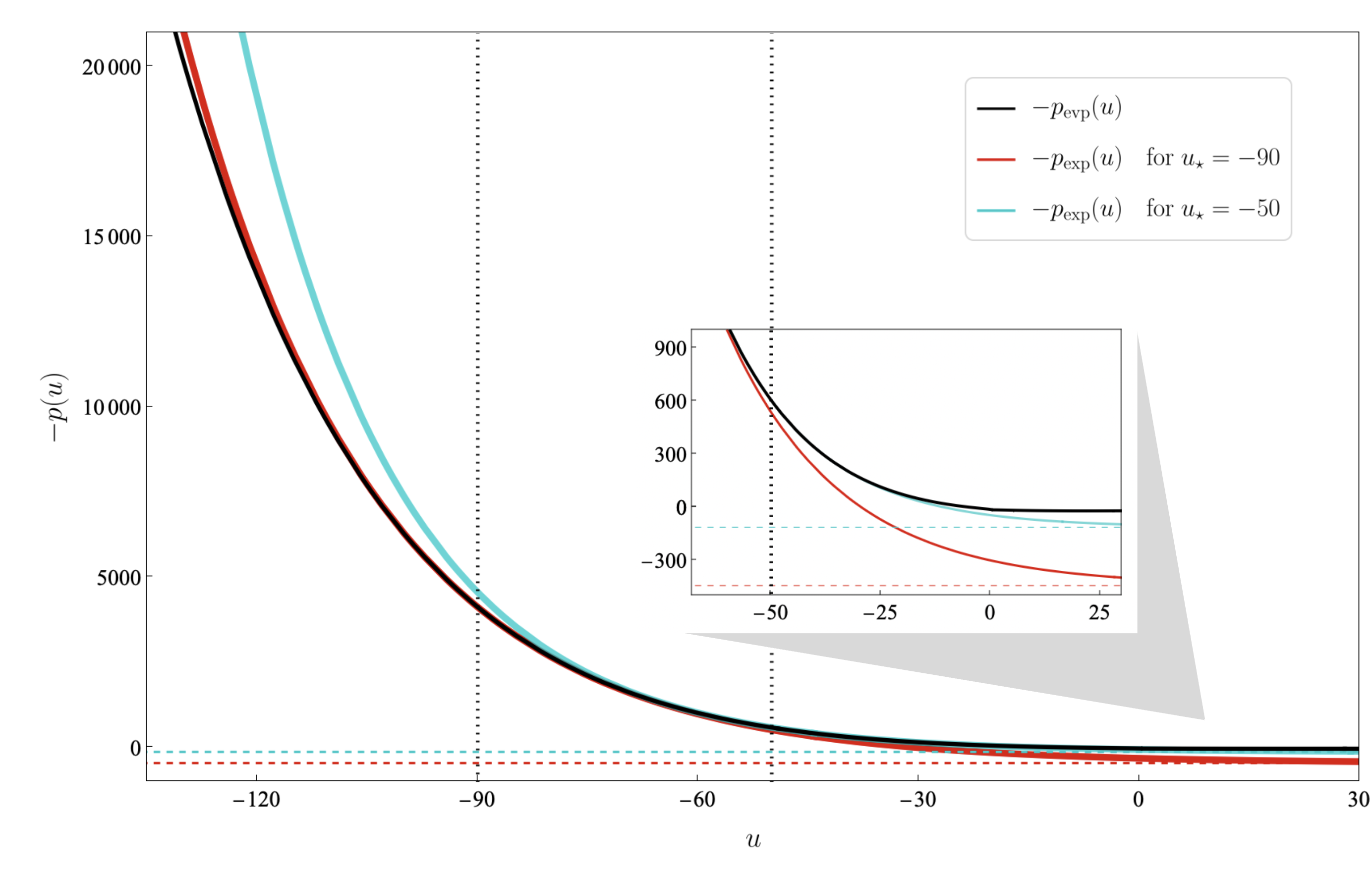}
    \vspace{-8pt}
    \caption{This figure compares the exact trajectory defined by $p_{\rm evp}(u)$ (black line) with the exponential approximation $p_{\rm exp}(u) = v_\star^{(H)} - 4 M_\star \, \dot{p}_\star \, 
e^{-\frac{u - u_\star}{4 M_\star}}$
for two different values of $u_\star$, namely $-90$ and $-50$ (in units of $M_0$). The constants are fixed to the values $\alpha =1$, $u_0 = 0$, $p_0 = 1$, and $\dot{p}_0 = 1$.
The figure shows that $p_{\rm exp}$ accurately describes the exact trajectory in an interval around $u_\star$.
The dashed lines correspond to the would-be horizons associated with the two chosen values of $u_\star$. The values of such are, respectively, $v_\star^{(H)} = -430.6$ and $v_\star^{(H)} = -101.8$.} 
    \label{expappro}
\end{figure}

Fig.~\ref{expappro}  illustrates the shape of the curve \eqref{pexpapprox} for two different values of $u_\star$ and compares them with the exact trajectory \eqref{pexact}. The figure shows that each curve approximates the exact trajectory only within a finite time interval. It also indicates the position of the instantaneous would-be horizon corresponding to each case.

When $M_\star$ is larger than a few Planck masses, the interval $\Delta u$ can be chosen large enough to accommodate wavelengths equal to or larger than the typical thermal wavelength $(\sim M_\star)$.  As $M(u)$ approaches the Planck scale, its rate of change becomes too rapid, and the interval $\Delta u$ within which Eq.~\eqref{pexpapprox} provides a good approximation to the actual trajectory becomes too small to define a temperature. This is the analogue of the stage at which the Hawking radiation emitted by a black hole ceases to be approximately thermal. We choose the instant $u_m$—the point at which the transition in Eqn.~\eqref{mirrorevp} to inertial motion begins—as an instant equal or earlier than the time when approximate thermality is lost.

 \subsubsection{Energy emitted during evaporation}

The energy flux arriving at $\mathcal{I}^{+}_R$ in the radiation zone $u \in [u_0, u_m)$ can be obtained by inserting
the function $p_{\rm evp}(u)$ into Eq.~\eqref{Tuu}. The result is
\begin{align}
    T_{uu} = \frac{1}{768\pi\, M^2(u)}\left[1 + \frac{8\alpha}{M^2(u)}\right],
\end{align}
valid for $u \in [u_0, u_m)$. The second term in this expression, and the time-dependence of $M$, capture the deviation from
thermality due to the slow ``evaporation'' of the mirror trajectory.

In the 1+1 dimensional model under study, the total energy emitted by the mirror in the radiation zone is obtained by integrating $T_{uu}$ from $u = u_0$ to
$u = u_m$:
\begin{align}\label{eevp}
    E_{\rm evp} = \frac{1}{768\pi\alpha}\left[M_0 - M(u_m)\right]
    - \frac{1}{96\pi}\left[M_0^{-1} - M(u_m)^{-1}\right].
\end{align}
The second term in this equation is negligible for macroscopic values of $M_0$. 

In the remainder of this article, we set $\alpha = 1/(768\pi)$ in Planck units, so that $E_{\rm evp}$ is
approximately equal to the mass difference between the beginning and the end of the
evaporating regime, up to a small correction.

In an actual evaporating black hole, energy conservation would require that $T_{uu}$ be exactly equal to $-\dot{M}$. 
Such equality could be enforced in the mirror model by slightly modifying either the trajectory 
$p_{\rm evp}(u)$ or the mass-loss function $M(u)$. However, doing so would introduce unnecessary complications 
without altering the essential physics. For this reason, we neglect the small discrepancy between 
$T_{uu}$ and $-\dot{M}$, as it has no significant effect on our results. 
A mirror trajectory satisfying $T_{uu} = -\dot{M}$ is discussed e.g. in Ref.~\cite{Good:2022wpw}.

\section{\label{sec:6} Partner modes at $\mathcal{I}^-$}

Consider a positive-frequency wavepacket in $\mathcal{I}^{+}_R$ with mean frequency $\omega>0$, centered at time $u_{\star} \in [u_0, u_{m})$, and with a time spread equal to $2\pi \epsilon^{-1}$,
where $\epsilon$ is a positive real number \cite{Wald:1975}. We choose this spread to satisfy the lower bound $\epsilon^{-1} \gg M_{\star}$. 
This ensures that the time-interval containing the main support of the wavepacket is large enough to accommodate the typical wavelengths of the mirror's thermal radiation, which are of order $\sim M_\star$. 

On the other hand, it is also necessary to impose the upper bound $\epsilon^{-1} \ll  M_{\star}^{2}/\sqrt{\alpha}$.
This guarantees that the wavepacket remains negligible outside the interval around $u_\star$ where the exponential approximation \eqref{pexpapprox} for the mirror trajectory holds.

As discussed above, wavepackets do not have compact support, although their contribution outside their time spread becomes negligibly small and can therefore be ignored. 
One may phrase this approximation differently, by replacing these wavepackets with smooth functions of strict compact support, obtained by truncating the wavepacket once its tail becomes sufficiently small, and smoothly bringing its value to zero via an interpolating function \cite{Evaporation_2024}. 
The interpolation can always be tuned so that it contributes negligibly to the calculations below. 
A consequence of making the wave packets compactly supported is that they are no longer composed exclusively of positive-frequency modes on $\mathcal{I}^{+}_R$. But their negative-frequency component can be made arbitrarily small and safely neglected. Moreover, they can be normalized with respect to the Klein-Gordon product.

In the following, we denote by $W^{\rm{out}}_{\omega u_{\star}}(u)$
the compactly supported functions approximating our wavepackets as described above.  These wavepackets arrive at $\mathcal{I}^{+}_R$ in the radiation zone, and the modes they describe there are in a mixed state when the field is prepared in the $|\rm in\rangle$ vacuum. 
The partner mode that purifies this particle can be computed as follows.

Let us propagate the mode $W^{\rm{out}}_{\omega u_{\star}}$ back to $\mathcal{I}^{-}_R$. The result is the following initial datum:
\begin{align}\label{up}
    W^{\rm{up}}_{\omega u_{\star}}(v) = W^{\rm{out}}_{\omega u_{\star}}\!\left(p_{\rm evp}^{-1}(v)\right).
\end{align}
The function $W^{\rm{up}}_{\omega u_{\star}}(v)$ is the progenitor of the Hawking mode $W^{\rm{out}}_{\omega u_{\star}}$ (the notation ``out'' and ``up'', as well as the name of other modes introduced shortly, are chosen following common terminology used in black hole spacetimes---see e.g. \cite{Frolov:1998wf,Evaporation_2024,Agullo:2023pgp}).

Because the support of the mode is small compared to the interval where the exponential approximation holds, 
we can safely replace the exact expression for $p_{\rm evp}$---Eqn.~\eqref{pexact}---by its exponential approximation around $u_\star$ given in Eqn.~\eqref{pexpapprox}. 
The partner mode can then be computed following a strategy similar to that first developed in Ref.~\cite{Wald:1975}. 
Namely, one first defines a new mode by reflecting $W^{\rm{up}}_{\omega u_{\star}}$ about the instantaneous would-be horizon $v_\star^{(H)}$:
\begin{align}\label{dn}
    W^{\rm{dn}}_{\omega u_{\star}}(v) := \bar{W}^{\rm{up}}_{\omega u_{\star}}\!\left(2v_{\star}^{(H)} - v\right),
\end{align}
(recall that the bar denotes complex conjugation).  Because of the exponential form of the approximation to $p_{\rm evp}$, both $W^{\rm{up}}_{\omega u_{\star}}(v)$ and $W^{\rm{dn}}_{\omega u_{\star}}(v)$ take the form of Rindler modes (propagated to $\mathcal{I}^{-}_R$) supported in the left and right Rindler wedges, respectively---where $v_\star^{(H)}$ is the boundary between wedges in $\mathcal{I}^{-}_R$. 
Thus, by analogy with the Unruh effect, $W^{\rm{dn}}_{\omega u_{\star}}(v)$ can be identified as the partner mode of $W^{\rm{up}}_{\omega u_{\star}}(v)$.

That $W^{\rm{dn}}_{\omega u_{\star}}(v)$ is the partner, can be checked explicitly using the tools introduced in Subsection~\ref{partner}. 
Specifically, one notes that the following combinations:
\begin{align}\label{sqz}
     s_{\omega}\!\left(W^{\rm{up}}_{\omega u_{\star}} 
        + e^{-4\pi M_{\star}\omega}\, \bar{W}^{\rm{dn}}_{\omega u_{\star}}\right),\\[4pt]\label{sqz2}
     s_{\omega}\!\left(W^{\rm{dn}}_{\omega u_{\star}} 
        + e^{-4\pi M_{\star}\omega}\, \bar{W}^{\rm{up}}_{\omega u_{\star}}\right),
\end{align}
are composed (essentially) only of positive frequencies at $\mathcal{I}^{-}$ \cite{Wald:1975}, 
where $s_{\omega} = [1 - \exp(-8\pi M_{\star}\omega)]^{-1/2}$ is a normalization constant.\footnote{They are ``essentially'' composed only of positive frequencies---not exactly---because of the restriction to compactly supported wave packets. The small negative-frequency contribution is negligible.}
This can be verified by Fourier analysis. 
Consequently, these two linear combinations are eigenfunctions of $J_{\rm{in}}$, the complex structure associated with the $|\rm{in}\rangle$ vacuum, with eigenvalue $i$. 
Using this fact, it is not hard to check that applying expression~\eqref{partnerformula} to $W^{\rm{up}}_{\omega u_{\star}}$ yields $W^{\rm{dn}}_{\omega u_{\star}}$. 
The partner mode subsystem defined in this manner is unique, in the sense explained in Subsection~\ref{partner}.

Two additional points are worth emphasizing. On the one hand, in the case of the Davies--Fulling trajectory, the partner mode of any Hawking particle is obtained by reflecting the corresponding wave packet $W^{\rm{up}}_{\omega u_{\star}}$ across the line $v = v_H$, defined as the asymptote of the exponential relation in Eqn.~\eqref{tranjectoryFD}, namely $v=0$.  
In contrast, for the trajectory $p_{\rm evp}$ we are considering, the partner is obtained by reflection across the instantaneous would-be horizon $v_{\star}^{(H)}$, which {\em differs for each Hawking mode}, since $v_{\star}^{(H)}$ depends on the center $u_{\star}$ of that mode on $\mathcal{I}^+_R$ [see Eqn.~\eqref{vhinst}].  
Thus, the calculation of partners is significantly richer in the case of a trajectory that mimics evaporation.

The center and support in $\mathcal{I}^-_R$ of the partner mode of a Hawking particle that is centered around $u_{\star}$ in $\mathcal{I}^+_R$ can be determined as follows.  
The instant $u_{\star}$ on $\mathcal{I}^+_R$ corresponds to $v = p_{\star}$ on $\mathcal{I}^-_R$.  
Hence, the center of the partner mode, denoted by $v_{\star}^{(p)}$, corresponds to the reflection of $p_{\star}$ about $v_{\star}^{(H)}$, namely
\begin{align}
    v_{\star}^{(p)} = 2v_{\star}^{(H)} - p_{\star}.
\end{align}
Recalling that $v_{\star}^{(H)} = p_{\star} + 4M_{\star}\dot{p}_{\star}$, the center of the partner mode is explicitly given by
\begin{align}\label{vpexact}
    v_{\star}^{(p)} = p_{\star} + 8M_{\star}\dot{p}_{\star}.
\end{align}

Next, we compute the the support of the partner mode on $\mathcal{I}^-_R$. 
There, the size of the support of the partner mode $W^{\rm{dn}}_{\omega u_{\star}}$ is the same as that of $W^{\rm{up}}_{\omega u_{\star}}$ (the progenitor of the Hawking wave packet), since $W^{\rm{dn}}_{\omega u_{\star}}$ and $W^{\rm{up}}_{\omega u_{\star}}$ are related by a reflection about $v_\star^{(H)}$.  
These supports are exponentially smaller than the support of the corresponding Hawking mode on $\mathcal{I}^+_R$, due to the blue-shift the mode experiences during backward propagation from $\mathcal{I}^+_R$ to $\mathcal{I}^-_R$.  
Specifically, if $u_\star\pm\Delta u$ denotes the support of the Hawking mode on $\mathcal{I}^+_R$, the support of its partner on $\mathcal{I}^-_R$ is given by the interval $[v_{\star}^{(p)} - \Delta v_{\star}^{(-)},\, v_{\star}^{(p)} + \Delta v_{\star}^{(+)}]$, where $\Delta v_{\star}^{(\pm)}$ is 
\begin{align}\label{psupp}
\Delta v_{\star}^{(\pm)} \approx  \mp 4M_{\star}\dot{p}_{\star}\!\left(1 - e^{\pm \frac{\Delta u}{4M_{\star}}}\right).
\end{align}

Note that while the support of the Hawking mode on $\mathcal{I}^+_R$ is symmetric with respect to its center, on $\mathcal{I}^-_R$ the supports of both the progenitor and the partner are not symmetric.  
This asymmetry arises from the exponential $u$-dependence of the blue-shift, which affects different parts of the support unequally.  
Furthermore, $\Delta v_{\star}^{(\pm)}/\dot{p}_0$ is exponentially smaller than $\Delta u$, as can be seen by recalling that $\dot{p}_{\star}$ decreases exponentially fast with increasing $u_{\star}$ [see Eqn.~\eqref{pdotexact}].

\section{\label{sec:7} Partner modes at $\mathcal{I}^+$}

As already emphasized earlier, it is important to appreciate that the partner of a Hawking particle in the state $|\rm in \rangle$ is determined solely from the form of the mirror trajectory in the radiation zone. 
Because the initial value problem is well-posed, $W^{\rm{dn}}_{\omega u_{\star}}$ continues to purify the corresponding Hawking mode at any later time, regardless of the mirror’s future trajectory. 
On the contrary, the shape of the partner mode $W^{\rm{dn}}_{\omega u_{\star}}$ when it eventually reaches $\mathcal{I}^+_R$ does depend on that future trajectory. 
In particular, the amount of energy the partner carries to $\mathcal{I}^+_R$ is sensitive to the state of motion of the mirror at the instant of reflection. 
As already emphasized earlier, this illustrates that there is no obvious constraint on the energy cost of purification.
See Fig.~\ref{fig2} for an illustration of the reflection process.

We are interested in understanding the plausibility of purification at no energy cost. 
For this to occur, all partners must reflect off the mirror when its trajectory has already returned to inertial motion, $p_{\rm in}(u)$. 
This imposes a constraint on the duration of the transition region, which can be obtained analytically as follows.

It is easy to check from Eqn.~\eqref{psupp} that the lower end of the support of the partner mode, given by $v_{\star}^{(p)} - \Delta v_{\star}^{(-)}$, lies exponentially close to $v_{\star}^{(H)}$, since recall that $\Delta u\gg M_{\star}$ for our wavepackets. 
It follows that vacuum purification occurs only if $v_{\star}^{(H)}$ is larger than $p_{\rm tr}(u_f)$---the value of $v$ at which the transition ends. 
This yields the following restriction on the duration of the transition period, for all $u_{\star} \in [u_0, u_m)$:
\begin{align}\label{vpur}
\frac{p_{\rm tr}(u_{f}) - p_{\rm tr}(u_{m})}{\dot{p}_0} \leq  
4\, e^{-M_0^2/(8\alpha)} 
\bigg\lbrace 
M(u_m)\, e^{M^2(u_m)/(8\alpha)}
+ \sqrt{2\pi\alpha} 
\bigg[
\mathrm{erfi}\!\left(\frac{M_{\star}}{\sqrt{8\alpha}}\right)
- \mathrm{erfi}\!\left(\frac{M(u_m)}{\sqrt{8\alpha}}\right)
\bigg]
\bigg\rbrace .
\end{align}

When the partner reflects while the mirror is inertial, it will emerge at $\mathcal{I}^+_R$ as a Rindler  mode. 
The reflection will, however, shift the frequency content of the mode, either towards the blue or the red, depending on the final velocity of the mirror.

Two limiting cases are worth highlighting. 
If the mirror continues with a velocity similar to the one it had at the end of the evaporation period—i.e., if it does not experience a large deceleration during the transition—then the reflection of the partner mode on its way from $\mathcal{I}^-_R$ to $\mathcal{I}^+_R$ will red-shift it. 
This red-shift is significant: The resulting mode is red-shifted with respect to the original Hawking mode. This is so because $\dot{p}(u)$ decreases exponentially during evaporation, and thus its final value is smaller than the value it had at the time the Hawking mode was reflected on its way from $\mathcal{I}^+_R$ to $\mathcal{I}^-_R$ when propagated back in time. 

The other limit corresponds to the case in which the mirror decelerates and returns to a velocity  similar to the one it had  before the radiation period. 
In this case, the partner does not experience any further frequency shift upon reflection, and its support and frequency content at $\mathcal{I}^+_R$ is exponentially blue-shifted relative to the initial Hawking mode. 
Note, however, that this scenario requires fine-tuning of the mirror’s deceleration during the transition period so as to precisely cancel the net velocity gain during the radiation regime. 
Since that velocity depends on the initial value of $M_0$, the mirror’s trajectory during the transition must be adjusted accordingly. Furthermore, as discussed in the next section and anticipated by Wald \cite{Wald:2019ygd}, the large deceleration induces a large burst of energy radiated to $\mathcal{I}^+_R$.

\section{\label{sec:7} Energy cost of purification}

During the evaporation period, the energy reaching $\mathcal{I}^+_R$ in the form of Hawking radiation, $E_{\rm evp}$, was computed in Eqn.~\eqref{eevp}. Using $\alpha = 768\pi$ in Planck units, it is approximately given by $E_{\rm evp} \approx M_0 - M(u_m)$.

After the transition, the trajectory of the mirror becomes inertial and no further energy is emitted. Thus, the transition region is the only stage where potential issues with the energy budget could arise, for instance in the form of an energy burst. We say that the energy budget is consistent if the total energy emitted $E_{\rm tot}$ during the evaporation and transition periods combined satisfies $E_{\rm tot} \lesssim M_0$. 

For a smoth trajectory of the form 
\begin{align}\label{mm}
p(u)=
\begin{cases}
    p_{\mathrm{evp}}(u) & u_0\leq u < u_{m},\\[6pt]
    p_{\mathrm{tr}}(u), & u \in [u_m,u_f),\\[6pt]
    p_{\mathrm{in}}(u), & u \ge u_f,
\end{cases}
\end{align}
with $p_{\mathrm{evp}}(u)$ given in Eqn.~\eqref{pexact}, the total energy emitted, including both the radiation and the transition period,  can be computed by integrating the energy flux given in Eqn.~\eqref{Tuu} in the interval $[u_0,u_f]$:
\begin{align}\label{totalenergy}
E_{\rm tot} = M_0 - M(u_m) - \frac{1}{96\pi M_0} + \frac{1}{48\pi} \int_{u_m}^{u_f} du \, \left[ \frac{\ddot{p}_{\rm tr}(u)}{\dot{p}_{\rm tr}(u)} \right]^2 .
\end{align}
Since $M_0 \gg 1$ in Planck units, our model remains energetically consistent provided that the last integral in Eqn.~\eqref{totalenergy} (which is positive) satisfies $\lesssim M(u_m)$. This condition defines a precise restriction on the allowed transition trajectories $p_{\rm tr}(u)$.

Recall that, as discussed in Section \ref{sec:5}, the local redshift factor grows exponentially as the mass decreases during the evaporation period, until $\dot{p}_{\rm evp}(u)$ reaches its minimum value at $u=u_m$:
\[\dot{p}_{\rm evp}(u_m)=\dot{p}_0\exp\left[\frac{M^2(u_m)-M_0^2}{8\alpha}\right]~.\]
This factor implies a huge velocity of the mirror [see Eqn.~\eqref{velocitymirror}] at the beginning of the transition epoch, relative to its initial motion just before the evaporation stage. If one demands the mirror to become not only inertial but actually at rest with respect to this initial motion (determined by $\dot{p}_0$), a large deceleration would be required. Such a large deceleration would require either (i) large values of the derivative of $\dot p$, inducing a large energy burst in a very short lapse of time $u_f-u_m$; or (ii) a very long transition period with a slow deceleration. Either case would violate the energy budget constraint.

In other words, vacuum purification requires that $\dot{p}_{\rm in}$ during the purification stage not differ too much from its value at the end of the evaporation period, namely $\dot{p}_{\rm evp}(u_m)$. This final value is exponentially smaller than the initial value $\dot{p}_0$, by a factor of order $e^{-M_0^2/(8\alpha)}$—an extraordinarily small number for $M_0$ larger than the Planck mass. This represents an important physical constraint on purification of Hawking radiation by vacuum fluctuations.

We conclude this section by noting that, for the mirror trajectory we are considering,  given in Eq.~\eqref{mm}, it is possible to determine the support of the would-be partner of the partner modes.  
Appendix~\ref{sec:App} contains a quantitative characterization of this support, and of  when the would-be partner of the partner overlaps with some of the Hawking modes. The analysis in Appendix~\ref{sec:App} offers a complementary way to understand the upper bound on $\dot{p}_{\rm in}$ during the purification period required to avoid an unphysical energy burst. We have placed the calculation of the location of the would-be partner of the partner in an Appendix because this information is not actually required to evaluate the energy budget, which is unambiguously, and more transparently determined by the energy flux discussed above. Nevertheless, we have included such analysis in this article in view of the prominent role these modes have played in previous discussions~\cite{Wald:2019ygd,Osawa:2024fqb}.

\section{\label{sec:8} An energetically consistent analog of evaporation and purification by vacuum fluctuations}

As anticipated in Ref.~\cite{Wald:2019ygd} and quantitatively discussed in the previous section, having both a consistent energy budget and vacuum purification requires the final inertial state of motion of the mirror to be ``boosted'' relative to the trajectory the mirror had before evaporation, in the sense that the  quotient $\dot{p}_{0}/\dot{p}_{\rm in}$ is very large---in fact, extraordinarily large for $M_0$ significantly greater than one in Planck units. This implies that the inertial velocity of the mirror at late times is extraordinarily large compared to its velocity at the onset of the radiation period. 
{\em This is the trade-off for vacuum purification.} Whether such a situation could have an analog in actual black holes will be discussed in the next section.

The model discussed in this section  can be viewed as an extension of one of the models presented in Ref.~\cite{Wald:2019ygd}, with the extension consisting in that it possesses an analytical trajectory whose temperature increases in time, mimicking an actual evaporation process---rather than remaining time-independent.  

For the trajectory during the transition between evaporation and purification, we consider in this section the simplest possible case that yields a consistent energy budget. 
This corresponds to an instantaneous transition. This choice does not yield a smooth mirror trajectory---the second derivative of the trajectory is discontinuous. The purpose of this construction is to imitate a smooth but sufficiently short transition.

The form of the trajectory is
\begin{align}\label{mevap}
p(u)=
\begin{cases}
    p_{\mathrm{evp}}(u), & u_0\leq u < u_{f},\\[6pt]
    p_{\mathrm{in}}(u)=p_{\mathrm{evp}}(u_f)+\dot p_{\rm in}\, u, & u \ge u_f,
\end{cases}
\end{align}
with $p_{\mathrm{evp}}(u)$ given in Eqn.~\eqref{pexact}, and
\begin{align}\label{instantparams}
&p_{\mathrm{evp}}(u_f)=v_{\star}^{(H)} - 4\dot{p}_{\star}\, e^{-M_{\star}^2/(8\alpha)}\, \bigg\lbrace  M(u_f)\, e^{M^2(u_f)/(8\alpha)}+\sqrt{2\pi\alpha}\bigg[\mathrm{erfi}\left(\frac{M_{\star}}{\sqrt{8\alpha}}\right)-\mathrm{erfi}\left(\frac{M(u_f)}{{\sqrt{8\alpha}}}\right)\bigg]\bigg\rbrace ,\nonumber \\[4pt]
&\dot p_{\rm in}=\dot p_{\mathrm{evp}}(u_f)=\dot{p}_0\,\exp{\left[\frac{M^2(u_{f})-M_0^2}{8\alpha}\right]}.
\end{align}
We assume $M(u_f)\neq 0$. Notice that the last equation shows that, when $M_0^2 - M^2(u_{f}) \gtrsim 1$ in Planck units, $\dot p_{\rm in}\ll \dot{p}_0$. 
This implies that modes reflected at the mirror  after $u=u_f$ will experience a large redshift. 

For this trajectory, one can analytically verify that the position $v_*^{(H)}$ of the instantaneous would-be horizon for a Hawking mode arriving at $\mathcal{I}^+_R$ at the instant $u_*$ satisfies $v_*^{(H)}>p(u_f)$ for all $u_*$ in the radiation zone. From the analysis of Sec.~\ref{sec:6}, it then follows that all partner modes are supported in $\mathcal{I}^+_R$ for $u>u_f$, where the state is indistinguishable from the vacuum. Hence, for this mirror trajectory Hawking radiation is purified by vacuum fluctuations. 

Before analyzing the energy budget, it is instructive to compare two-point field correlations between the radiation and purification zones in the $|\rm in\rangle$ and $|\rm out\rangle$ vacua. Let $u_*$ denote an instant in the radiation zone of $\mathcal{I}^+_R$, and let $u_*^{(p)}$ denote the center of its partner mode in the purification zone. From Sec.~\ref{sec:6} we obtain:
\be 
u_*^{(p)}=u_*+4M_{\star}\exp\left[\frac{M^2_{\star}-M^2(u_{f})}{8\alpha}\right]\,+4M(u_f)+\sqrt{2\pi\alpha}\bigg[\mathrm{erfi}\left(\frac{M_{\star}}{\sqrt{8\alpha}}\right)-\mathrm{erfi}\left(\frac{M(u_f)}{{\sqrt{8\alpha}}}\right)\bigg] .
\ee 
With this, for $M_{\star}\gg M(u_f)$ and larger than one in Planck units, we can estimate the ratio of correlations:
\be \label{corrinvsout}
\frac{\langle {\rm in}|\partial_u\varphi(u_*^{(p)})\partial_u\varphi(u_*)|{\rm in}\rangle}
{\langle {\rm out}|\partial_u\varphi(u_*^{(p)})\partial_u\varphi(u_*)|{\rm out}\rangle}
=\frac{\dot p(u_*^{(p)})\dot p(u_*)\, (u_*^{(p)}-u_*)^2}{[p(u_*^{(p)})-p(u_*)]^2}
\approx \frac{1}{4}e^{M_*^2/(8\alpha)} \gg 1,
\ee
This calculation reveals that the correlations between $u_*$ and $u_*^{(p)}$ are much stronger in the $|\rm in\rangle$ vacuum than what they would be in the $|\rm out\rangle$ state. These are the correlations that purify the Hawking radiation. 

Figure~\ref{mirror} illustrates some of the key aspects of the purification process for the mirror trajectory given in Eqn.~\eqref{mevap}. A few points are worth emphasizing:

\begin{figure}
\centering   
    \includegraphics[width=.6\linewidth]{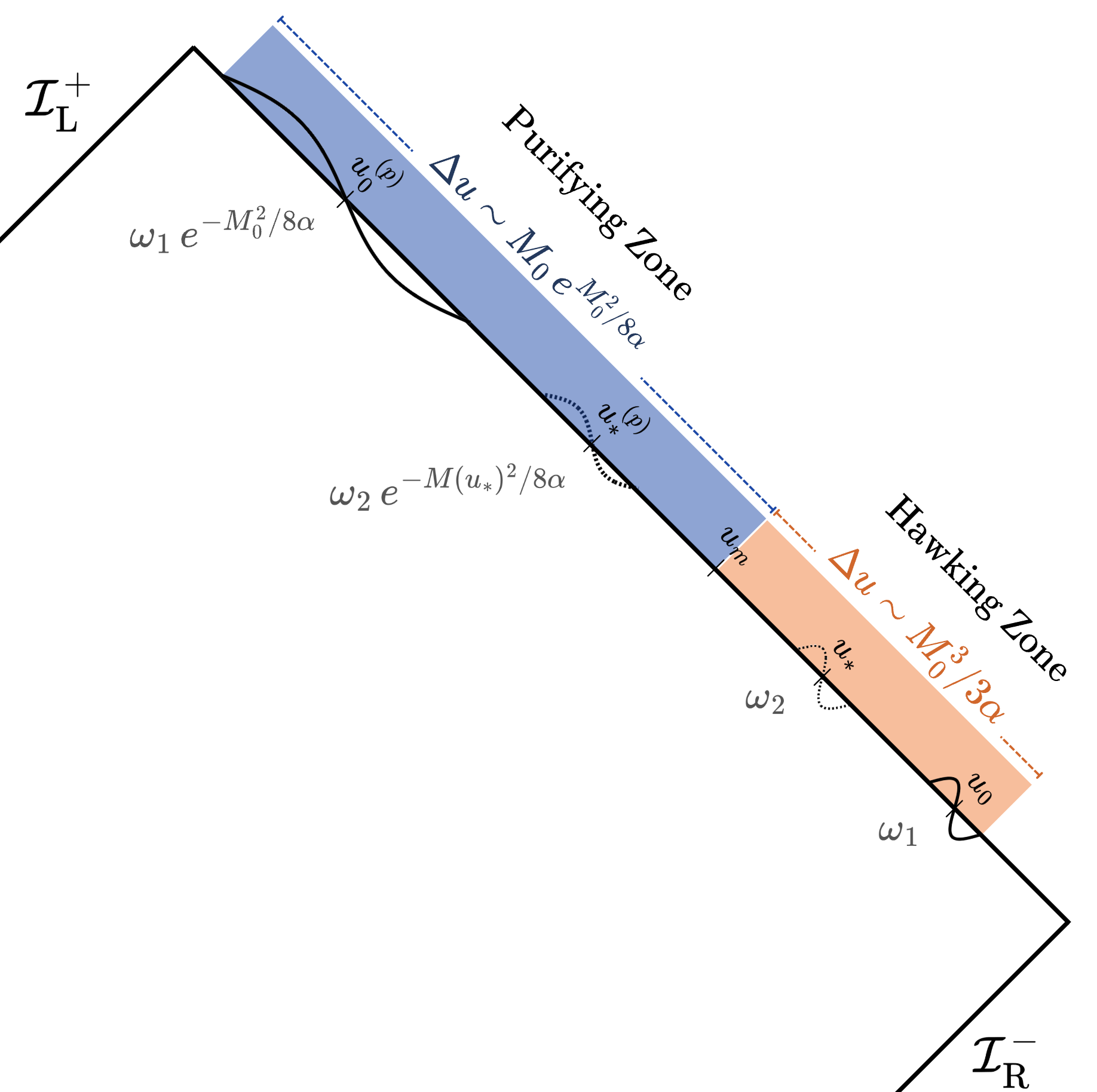}
    \vspace{-8pt}
    \caption{Depiction of the purification of Hawking radiation by late-time vacuum fluctuations. The figure illustrates points 1, 2, and 3 described in the text above.}
    \vspace{8pt}
    \label{mirror}
\end{figure}

\begin{enumerate}
\item The duration of the purification zone is (at least) of order $\sim M_0\, e^{M_0^2/(8\alpha)}$. This duration grows {\em exponentially} with $M^2_0$, while the duration of the radiation zone, 
$\sim \frac{M_0^3}{3\alpha}$, grows polynomically as $M_0^3$. 

\item The first Hawking modes emitted (around $u_0$) are the last ones to be purified. This is, the order of arrival of the Hawking modes is the opposite of the order of arrival of the respective partners. 

\item If $\omega$ is the central frequency of a Hawking mode arriving at $\mathcal{I}^+_R$ at $u_*$, the central frequency of its partner is of order 
$\omega\, \dot p(u_*^{(p)})/\dot p(u_*)\approx \omega \, e^{-M_\star^2/(8\alpha)}$. 
This implies, on the one hand, that the partners are {\em exponentially redshifted} relative to the original Hawking modes. On the other hand, since $M_\star$ decreases throughout the evaporation period, the redshift becomes more and more pronounced as time advances: modes in the purification zone are increasingly infrared---their frequencies decrease exponentially while their support extends over longer intervals (see Fig.~\ref{mirror}). 
\end{enumerate}

To conclude, we turn our attention to the energy flux arriving at $\mathcal{I}^+_R$ as predicted by this model. As mentioned above, the trajectory resulting from an instantaneous transition displays a discontinuity in its second derivative at $u=u_f$. This implies that the energy flux acquires a Dirac delta contribution. Nevertheless, $T_{uu}(u)$ is well defined as a distribution on $\mathcal{I}^+_R$, yielding a finite value for the integrated energy emitted over any finite interval. Explicitly, for $u>u_0$ we obtain
\begin{align}
T_{uu}(u)=\frac{1}{768\pi M^2(u)}\left[1+\frac{4\alpha}{M^2(u)}\right]\Theta(u_f-u)
-\frac{1}{96\pi M(u_f)}\,\delta(u_f-u)~,
\end{align}
where $\Theta(x)$ denotes the Heaviside step function [$\Theta(x)=1$ for $x>0$ and $0$ otherwise]. The first term represents the energy flux emitted in the radiation zone, corresponding to approximately thermal emission, while the second term arises from the instantaneous transition. Note that the latter is negative—this is consistent with the analysis in Ref.~\cite{Bianchi:2014vea}, where it was argued that the emission of negative energy is unavoidable for any mirror trajectory leading to a unitary evolution of the field from $\mathcal{I}^-_R$ to $\mathcal{I}^+_R$.

The integral of this energy flux is %
\begin{align}
E_{\rm evp}=\frac{1}{768\pi\alpha}\left[M_0-M(u_f)\right]
-\frac{1}{96\pi}\left[M_0^{-1}-M(u_f)^{-1}\right],
\end{align}
where $\alpha=1/(768\pi)$. For $M(u_f)\gtrsim 1$ in Planck units, the total energy emitted is smaller than $M_0$.

In this model, one can explicitly verify that the would-be partners of the partners overlap with the Hawking modes themselves (see Appendix \ref{sec:App}). This shows that, for our trajectory, the inertial particles predicted in Ref.~\cite{Wald:2019ygd}—created as an indirect consequence of purification via late-time vacuum fluctuations—are not independent from the Hawking quanta themselves. Therefore, no additional energy cost is implied by the purification process.
 
\section{\label{sec:9} Discussion}

This article investigates quantitatively the possibility that Hawking-like quanta emitted by a moving mirror can be purified by late-time vacuum fluctuations. Our motivation comes from recent discussions in Refs.~\cite{Wald:2019ygd,Osawa:2024fqb} concerning whether vacuum purification entails an enormous indirect energy cost, which would render such mechanisms for restoring unitarity inapplicable to realistic black-hole scenarios.

Our analysis includes a formulation of the problem in terms of local correlation functions and energy fluxes, avoiding any reliance on a particle interpretation of quantum field theory or on additional approximations.

We confirm that vacuum purification, as proposed in Ref.~\cite{H_S_U_2015}, is indeed possible. Local field excitations in the radiation zone of $\mathcal{I}^+_R$ are described by a thermal two-point correlation function. At late times in $\mathcal{I}^+_R$, in the region we refer to as the purification zone, field correlations become identical to those of the natural vacuum state. Nevertheless, there exist strong correlations (and entanglement) between field excitations in the radiation and purification zones, which ensure that the total state of the field on $\mathcal{I}^+_R$ remains pure.

We introduce an analytic mirror trajectory $\gamma(u) = (u, p(u))$ that produces Hawking radiation in the same manner as an evaporating black hole, namely with a temperature that increases adiabatically as energy is radiated to $\mathcal{I}^+_R$, and have quantitatively analyzed aspects of the purification process, including total energy radiated and location and support of the partner modes.

We find that vacuum purification imposes stringent constraints on the mirror trajectory in the transition region between radiation and purification. In particular, the length of this region in $\mathcal{I}^{-}_R$ is bounded from above by a very small quantity, given explicitly in Eq.~\eqref{vpur}. Moreover, the enormous velocity accumulated by the end of evaporation implies that the mirror cannot return to its initial state of motion during the transition period; otherwise, it would emit a prohibitively large burst of radiation. Altogether, we conclude that vacuum purification requires the velocity of the mirror after the evaporation process to be exponentially large relative to its initial state of motion.

This implies that (relative to the Bondi frame on $\mathcal{I}^+_R$ defined by the coordinate $u$) there remains a residual, exponentially large redshift for waves propagating from $\mathcal{I}^-_R$ to $\mathcal{I}^+_R$ after the mirror returns to inertial motion. One can always define a new Bondi frame $u'$ in which this redshift is absent at late times; such a frame is related to the early-time frame $u$ by a large boost. In this sense, the indirect ``cost'' of vacuum purification is a boost between the early- and late-time Bondi frames for which reflection at the mirror produces no red- or blue-shift. This boost depends on the initial mass $M_0$ in an exponential manner.

The key question is whether such a residual boost could have an analogue in actual black holes. We emphasize that, in order for such a residual boost to disappear,  the internal evolution of the small, presumably Planck-size black hole would need to keep memory of the value of black hole mass $M_0$ before the evaporation started, in order to precisely cancel the boost between  Bondi frames at $\mathcal{I}^-$ and $\mathcal{I}^+$ caused by Hawking radiation during the long evaporation history. It is far from obvious how such fine-tune  cancellation could happen. 

Interestingly, a picture closely related to vacuum purification has being proposed within the loop quantum gravity community, using completely different arguments (see e.g. \cite{Ashtekar:2025ptw,Rovelli:2024sjl}). In this framework, the partners of Hawking modes are not exactly vacuum fluctuations, but extremely infrared modes carrying very small energy. Near-Planckian mass black holes can presumably emit such ultra–infrared modes at the end of the evaporation process because, despite their small masses and surface areas, they possess enormous interior volumes \cite{Christodoulou:2014yia}. This, in turn, makes it possible to purify all Hawking quanta using only the small amount of energy remaining in the system.

In our view, a key message from the analysis of moving mirrors is that there is no sharp relationship between purification and energy---yet such a relation has been implicitly assumed in many discussions of black-hole evaporation. The possibility of purifying Hawking radiation with a small energy cost, compatible with the energy budget at the end of the evaporation process, is a viable idea which could open new avenues for understanding the fate of black holes, and merit further attention.

\acknowledgments
The content of this paper has benefited from discussions with A. Ashtekar, E.Bianchi, A. Delhom, C. Rovelli, and T. Thiemann. I.A. supported by the NSF grants PHY-2409402 and PHY-2110273, by the RCS program of Louisiana
Boards of Regents through the grant LEQSF(2023-25)-RD-A-
04,  by the Hearne Institute for Theoretical Physics and by Perimeter Institute of Theoretical Physics through the Visitor fellow program. Research at Perimeter Institute is supported
in part by the Government of Canada through the Department of Innovation, Science and Industry Canada
and by the Province of Ontario through the Ministry of
Colleges and Universities. B.E.N. acknowledges partial support by Project No. PID2023-149018NB-C41 from Spain. This work was made possible through the support of the Enrico Fermi Fellowships led by the Center for Spacetime and
the Quantum, and supported by Grant ID 63132 from the John Templeton Foundation. 
This work was made possible through the support of the WOST (WithOut SpaceTime) project (https://withoutspacetime.org), led by the Center for Spacetime and the Quantum (CSTQ), and supported by Grant ID 63683 from the John Templeton Foundation (JTF). The opinions expressed in this
publication are those of the authors and do not necessarily 
those of the Center for Spacetime and the Quantum.

\appendix

\section{\label{sec:App} Support of the would-be partners of the partners}

This Appendix contains a detailed analysis of the support at $\mathcal{I}^{+}_R$ of the partners of Hawking quanta, together with the field modes these partners  would be entangled with if the state was the $|\rm{out}\rangle$ vacuum. We focus on the  physically interesting  case in which radiation from the evaporation period is purified by vacuum fluctuations.

If the partner modes bounce off our mirror when it follows an inertial trajectory, their evolution to $\mathcal{I}^{+}_R$
can be computed by simply substituting the linear relation $v=p_{\rm{in}}(u)$, for $u\geq u_f$, in their initial data $W^{\rm{dn}}_{\omega u_{\star}}(v)$ at $\mathcal{I}^{-}_R$. Recalling that
such data is related to the Hawking modes via Eqs.~\eqref{up} and \eqref{dn}, it follows that the partners display the following (approximate) form at $\mathcal{I}^{+}_R$:
\begin{align}
W^{\rm{p}}_{\omega u_{\star}}(u)=\bar{W}^{\rm{out}}_{\omega u_{\star}}(l_{\star}(u))~,
\end{align}
where we have defined the following logarithmic function:
\begin{align}\label{lstar}
l_{\star}(u)=u_{\star}-4M_{\star}\ln{\left[\frac{p_{\rm{tr}}(u_f)-v_{\star}^{(H)}+\dot{p}_{\rm{tr}}(u_f)(u-u_f)}{4M_{\star}\dot{p}_{\star}}\right]}\,.
\end{align}

Using Eq.~\eqref{psupp}, the support of each of these partners at $\mathcal{I}^{+}_R$ can be computed to be approximately equal to the interval
$[u_{\star}^{(-)},u_{\star}^{(+)}]$, with
\begin{align}\label{upm}
u_{\star}^{(\pm)}=u_f +\dot{p}_{\rm{tr}}(u_f)^{-1}\left[\, v^{(H)}_{\star}-p_{\rm{tr}}(u_f)+4M_{\star}\dot{p}_{\star}e^{\pm \frac{\Delta u}{4M_{\star}}}\right]~,
\end{align}
where we recall that $u_{\star}\pm \Delta u$ denotes the support of the Hawking modes at $\mathcal{I}_R^+$.

The logarithmic dependence of the partner modes at $\mathcal{I}^{+}_R$
when they bounce off the inertial part of the mirror reveals that they 
are (approximate) left-Rindler modes there. The corresponding right-Rindler modes,  which they would be entangled with if the state was the $|\rm{out}\rangle$ vacuum---i.e., the would-be partners of the partners---are obtained
via reflection across the value of $u$ that makes the argument of the logarithm in \eqref{lstar} to vanish \cite{Wald:2019ygd}. The reader can easily check that this instant is simply
the image of the instantaneous would-be horizon under the linear mirror trajectory, namely
\begin{align}
u_{\star}^{(H)}=u_f +\dot{p}_{\rm {tr}}(u_f)^{-1}\left[v_{\star}^{(H)}-p_{\rm{tr}}(u_f)\right].
\end{align}

The support of the would-be partner of the partners  is obtained via reflection of the interval $[u_{\star}^{(-)},u_{\star}^{(+)}]$ across $u=u_{\star}^{(H)}$, namely
$[2u_{\star}^{(H)}-u_{\star}^{(+)},2u_{\star}^{(H)}-u_{\star}^{(-)}]$.  If this support overlaps with that of the Hawking modes, the two types of modes are not orthogonal with respect to the Klein-Gordon product. In the framework of Quantum Field Theory, this means that they do not define independent notions of particles.

Whether the supports of the would-be-partners of partners and Hawking modes overlap depends on the form of the mirror's trajectory during the transition period,  $v=p_{\rm{tr}}(u)$. Indeed, we can explicitly evaluate the past end of their support, $u_*^{(+)}$, using Eqs.~\eqref{pdotexact} and \eqref{vhinst} to explicitly write $v_{\star}^{(H)}$ and $\dot{p}_{\star}$ in terms of $\dot{p}_0$, $M_{\star}$, and $M_0$: 
\begin{align}
2u_{\star}^{(H)}-u_{\star}^{(+)}=u_f+\frac{\dot{p}_0}{\dot{p}_{\rm{tr}}(u_f)}\Bigg[&\frac{p_{\star}-p_{\rm{tr}}(u_{f})}{\dot{p}_0}+4M_{\star}\exp\Bigg(\frac{M_{\star}^2-M_0^2}{8\alpha}\Bigg)\left(1-e^{\frac{\Delta u}{4M_{\star}}}\right)\Bigg]~.
\end{align}
Two observations can be drawn from this expression:
\begin{enumerate}
    \item The would-be-partners of partners always have support to the past of $u_f$ (the end of the transition region).
    \item They can only lie to the past of $u_m$ (the end of the radiation region), and therefore overlap with the Hawking modes, if the global change in the mirror's velocity, which is determined by $\dot{p}_0/\dot{p}_{\rm{tr}}(u_f)$, is large.
\end{enumerate}
The reader can check that these observations are nontrivially based on: (i) The behavior of the function $p_{\rm{evp}}$ (see Sec.~\ref{sec:5}), (ii) the smallness of $p_{\rm{tr}}(u_f)-p_{\rm{tr}}(u_m)$ relative to $\dot{p}_0$ Planck seconds [see Eqn.~\eqref{vpur}], and (iii) the fact that $M_{\star}\ll \Delta u\ll M_{\star}^2/\sqrt{\alpha}$.

The second observation reinforces the restriction that the redshift factor should not change too much during the transition [recall that $\dot{p}(u_m)\ll \dot{p}_0$]. Otherwise, there would exist non-overlapping would-be-partners of partners carrying energy to $\mathcal{I}^{+}_R$ independently from the Hawking flux. Let us, then, focus on the illustrative case of the instantaneous transition introduced in Sec.~\ref{sec:8} [see Eqn.~\eqref{mevap}]. The past end of the support of the would-be-partners of partners in this scenario is
\begin{align}
2u_{\star}^{(H)}-u_{\star}^{(+)}=&u_f+4M(u_f)+4\sqrt{2\pi\alpha}\,e^{-M_{\star}^2/(8\alpha)}\Bigg[\mathrm{erfi}\left(\frac{M_{\star}}{\sqrt{8\alpha}}\right)-\mathrm{erfi}\left(\frac{M(u_f)}{{\sqrt{8\alpha}}}\right)\Bigg]\nonumber \\&-4M_{\star}\exp\Bigg[\frac{M_{\star}^2-M^2(u_f)}{8\alpha}+\frac{\Delta u}{4M_{\star}}\Bigg]~.
\end{align}
This quantity is clearly smaller than $u_f$ for $M_{\star}=M(u_f)$. Moreover, its derivative with respect to $M_{\star}$ is strictly negative for all $M_{\star}\geq M(u_f)\gtrsim 1$, since the Hawking modes are defined so that  $\Delta u\ll M_{\star}^2 /\sqrt{\alpha}$. It then follows that $2u_{\star}^{(H)}-u_{\star}^{(+)}<u_f$ for all $u_{\star}$, so all the would-be-partners of partners overlap with the Hawking modes and, therefore, cannot be viewed as independent excitations.

\bibliography{library}

\end{document}